\documentclass[reprint,twocolumn,superscriptaddress, amssymb,aps,pra]{revtex4-1}
\usepackage{graphicx}
\usepackage{xcolor} 
\usepackage{amsmath}
\usepackage{hyperref}

\begin{document}
\title{Demonstration of Erasure Conversion in a Molecular Tweezer Array}

\author{Connor M. Holland}
\affiliation{Department of Physics, Princeton University, Princeton, New Jersey 08544 USA}
\author{Yukai Lu}
\affiliation{Department of Physics, Princeton University, Princeton, New Jersey 08544 USA}
\affiliation{Department of Electrical and Computer Engineering, Princeton University, Princeton, New Jersey 08544 USA}
\author{Samuel J. Li}
\affiliation{Department of Physics, Princeton University, Princeton, New Jersey 08544 USA}
\author{Callum L. Welsh}
\affiliation{Department of Physics, Princeton University, Princeton, New Jersey 08544 USA}

\author{Lawrence W. Cheuk}
\email{lcheuk@princeton.edu}
\affiliation{Department of Physics, Princeton University, Princeton, New Jersey 08544 USA}

\date{\today}

\begin{abstract}
Programmable optical tweezer arrays of molecules are an emerging platform for quantum simulation and quantum information science. For these applications, reducing and mitigating errors that arise during initial state preparation and subsequent evolution remain major challenges. In this paper, we present work on site-resolved detection of internal state errors and quantum erasures, which are qubit errors with known locations. First, using a new site-resolved detection scheme, we demonstrate robust and enhanced tweezer array preparation fidelities. This enables creating molecular arrays with low defect rates, opening the door to high-fidelity simulation of quantum many-body systems. Second, for the first time in molecules, we demonstrate mid-circuit detection of erasures using a composite detection scheme that minimally affects error-free qubits. We also demonstrate mid-circuit conversion of blackbody-induced errors into detectable erasures. Our demonstration of erasure conversion, which has been shown to significantly reduce overheads for fault-tolerant quantum error correction, could be useful for quantum information processing in molecular tweezer arrays.
\end{abstract}

\maketitle

\section{Introduction}
Programmable optical tweezer arrays of ultracold molecules are an emerging platform for quantum science~\cite{kaufman2021quantum}. They combine the microscopic detection and control capabilities offered by programmable optical tweezer arrays~\cite{Endres2016atomarray,Labuhn2016atomarray} with useful properties of molecules such as rich internal structures and intermolecular electric dipolar interactions between long-lived states~\cite{yan2013observation,christakis2022probing,Gregory2024second}. In particular, the internal structure of molecules provides additional ways to encode quantum information and can be exploited for precision measurement experiments~\cite{Carr2009review,Bohn2017molreview,Safronova2018NewPhysics,Kozyryev2017PolyBSM,Anderegg2023trapCaOH}, while the intermolecular electric dipolar interaction is the key ingredient for high-fidelity quantum gates~\cite{demille2002quantum,Ni2018gate,Yu2019symtop} and simulation of novel quantum many-body Hamiltonians~\cite{Micheli2006spintoolbox,Carr2009review,Gadway2016StronglyInteracting,Bohn2017molreview,kaufman2021quantum}.

Starting with the first demonstrations of trapping and high-fidelity detection of individual molecules in optical tweezer traps~\cite{Anderegg2019Tweezer,Zhang2021NaCsArray,Holland2023bichromatic,Ruttley2023RbCsTweezer,Vilas2023CaOHTweezer}, recent work has demonstrated highly coherent rotational qubits in tweezers~\cite{burchesky2021rotcoh,Park2023rotcoh,Gregory2024second}, observed coherent dipolar interactions between individually held molecules, and demonstrated on-demand entanglement and 2-qubit gates between rotational molecular qubits~\cite{Holland2023Entanglement,Bao2023Entanglement}. These works establish the building blocks for quantum information processing and quantum simulation with molecular tweezer arrays. 

A major experimental challenge in any quantum science platform is reducing and mitigating errors that occur during the initial preparation of a quantum system and its subsequent evolution. In molecular tweezer arrays, one desires to deterministically prepare tweezer sites such that they are occupied by molecules from a single internal state. This preparation of molecular tweezer sites is challenging, and has been a bottleneck for applications in quantum simulation and information processing. In particular, the complex internal structure of molecules combined with substantial light shifts inside tweezer traps render high-fidelity optical pumping difficult~\cite{Lu2024RSC,Bao2023RSC}, thereby limiting achievable internal state purities. Concerning errors that can arise during quantum evolution, Raman scattering of tweezer trapping light or blackbody radiation can excite molecules out of the desired quantum states~\cite{Holland2023bichromatic}. While directly reducing these errors through better control is an important research area that has attracted much interest, a complementary approach is to mitigate their effects via measurement and feedback. In particular, for programmable tweezer arrays, because local control is available, site-resolved error detection could in some cases enable their subsequent removal.

Site-resolved detection of preparation errors has been used in atomic tweezer arrays to increase internal state preparation fidelities~\cite{Scholl2023Erasure}, increase robustness to radiative decays of high-lying Rydberg states~\cite{Scholl2023Erasure,Ma2023Erasure}, and to lower motional temperatures~\cite{Scholl2023HyperEntanglement}. 
In recent work with molecules that are assembled from two atoms and trapped in tweezer arrays, this has been used to enhance molecular formation efficiency and to reduce defects in molecular arrays~\cite{Picard2024siteselective,Ruttley2024feedback}.

Restricting to the specific case where information is stored in an array, an information error whose location is known is called an erasure. For qubits, such errors are called quantum erasures~\cite{Grassl1997erasurecodes,Bennett1997ErasureCapacity}. Recently, it has been pointed out that leakage errors, where a particle hosting a qubit exits its computational space, can be converted and detected as an erasure~\cite{Wu2022Erasure}. Notably, erasure conversion substantially raises fault-tolerant quantum error correction thresholds to practically achievable levels~\cite{Wu2022Erasure}. This important insight has led to several proposed erasure conversion schemes across a variety of qubit platforms~\cite{Wu2022Erasure, Kubica2023erasureproposal,Kang2023erasureproposal,Teoh2023erasureproposal}. Recently, the first demonstrations of quantum erasure conversion were achieved in atomic tweezer arrays of alkaline-earth atoms ~\cite{Ma2023Erasure,Scholl2023Erasure,Scholl2023HyperEntanglement} and superconducting circuits~\cite{Chou2023Erasure,Levine2024Erasure}.

In this paper, we present new site-resolved detection schemes to mitigate both state preparation errors and qubit leakage errors for laser-cooled molecular arrays. First, we demonstrate a new site-resolved detection scheme that enables robust and enhanced fidelities of preparing a tweezer site with a molecule in a single internal state. We achieve a record-level tweezer site preparation fidelity of $95.2(3)\%$, with internal state purity of $99.5(1)\%$, substantially surpassing the previous best reported fidelities of $\approx 80\%$~\cite{Holland2023Entanglement,Lu2024RSC}. Second, using a new composite erasure detection scheme, we demonstrate quantum erasure conversion in molecules for the first time. By utilizing a new hyperfine qubit encoding that is both highly coherent and compatible with a previously demonstrated 2-qubit gate~\cite{Holland2023Entanglement,Bao2023Entanglement}, we demonstrate a mid-circuit detection scheme that minimally affects qubit population and coherence. We further show mid-circuit erasure conversion of leakage errors caused by blackbody radiation, achieving improved qubit lifetimes and coherence times. 

Our new preparation scheme overcomes a major challenge in molecular tweezer arrays and opens the door to simulation of quantum many-body spin models with low defect rates. Our demonstration of quantum erasure conversion adds an important capability to the molecular quantum information processing toolkit, and opens the door to mid-circuit quantum error correction in molecular tweezer arrays.

\section{Framework for Site-Resolved Error Detection}
\label{sec:framework}
\begin{figure}[t]
	{\includegraphics[width=\columnwidth]{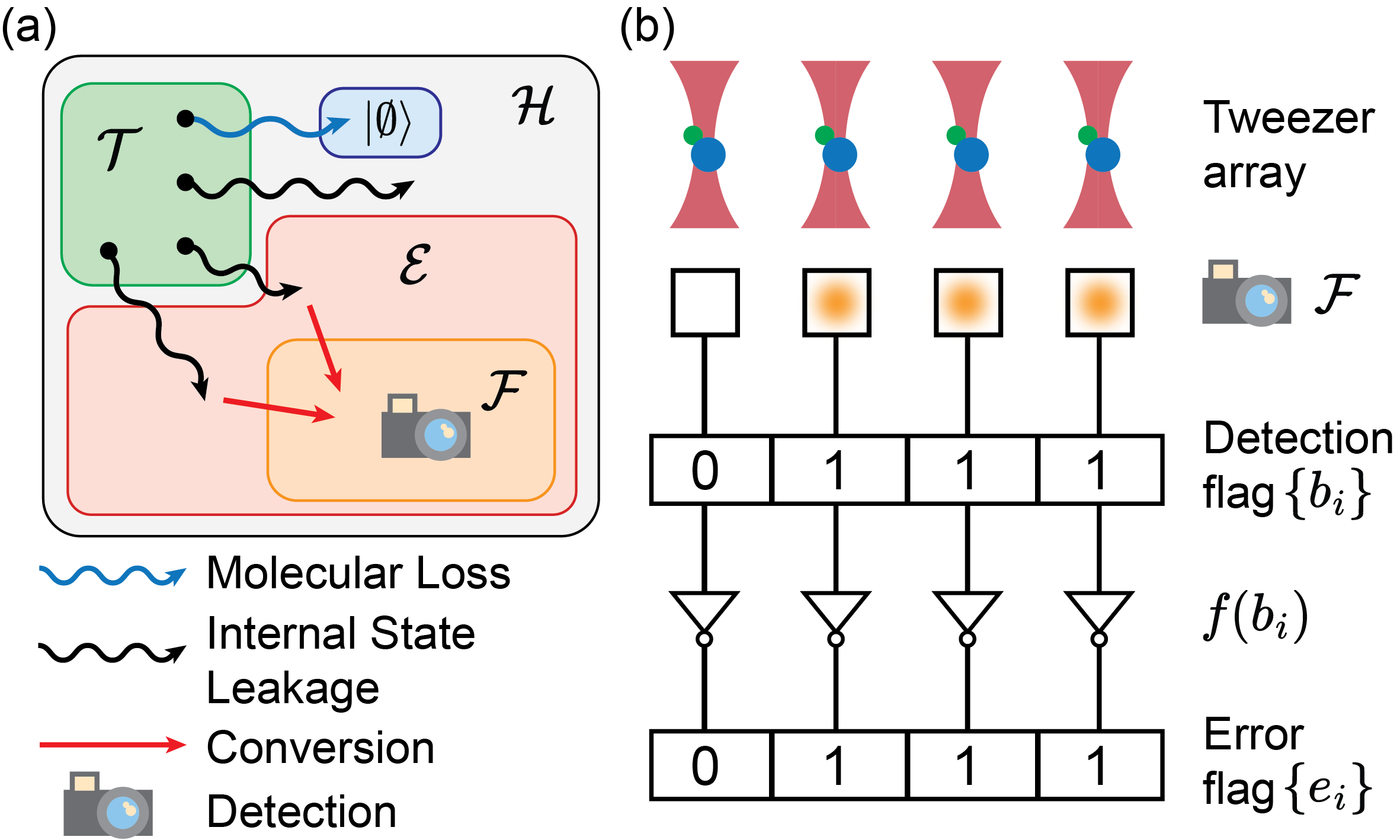}}
	\caption{\label{Fig_0} Framework for Site-Resolved Error Detection. (a) $\mathcal{H}$ is the Hilbert space on a tweezer site; $\mathcal{T}$ is the target space; $\mathcal{F}$, the detection manifold, consists of detectable states used to flag errors. Leakage errors transfer population from $\mathcal{T}$ to $\mathcal{E}$, which can subsequently be converted to $\mathcal{F}$ for detection. Loss of a particle results in an empty tweezer ($|\emptyset\rangle$). (b) Detection of total population in $\mathcal{F}$ is binarized into a bitstring $\{b_i\}$, from which the error flag $\{e_i\}$ is obtained.}
	\vspace{-0.2in}
\end{figure}

In atomic and molecular tweezer array experiments, each tweezer site is either empty or occupied by one particle. The local Hilbert space $\mathcal{H}$ at each site thus consists of $\left|\emptyset\right\rangle$, the state corresponding to an empty site, and the set of all internal states (Fig.~\ref{Fig_0}(a)). (We will ignore motional states of the particle within a tweezer here, although the same framework used here applies~\cite{Scholl2023HyperEntanglement}.) For a particular application, only a target subset of these states $\mathcal{T}$ is used. For example, in order to encode a qubit, at least two internal states are needed.

For site-resolved error detection, one uses a set of detectable internal states $\mathcal{F}$ to flag the presence of an error. The detection outcome is binarized and converted into an error flag $\{e_i\}$ where $e_i=1$ ($e_i=0$) indicates an error (error-free site)~(Fig.~\ref{Fig_0}(b)). $\{e_i\}$ could be used to remove errors mid-sequence, or to post-select for error-free instances by excising error-flagged data.

When at least one bit of information is encoded in $\mathcal{T}$, errors with a known location are called erasures~\cite{Grassl1997erasurecodes,Bennett1997ErasureCapacity}. In particular, population leakage out of $\mathcal{T}$ into a set of disjoint internal states $\mathcal{E}$ can be transferred to $\mathcal{F}$ and subsequently detected, a process known as erasure conversion~\cite{Wu2022Erasure}.

For site-resolved error detection to be practically useful, it must not affect error-free sites. Specifically, for quantum erasure detection, \textit{all populations and coherences} in $\mathcal{T}$ should be minimally affected by detection. We will reserve the term mid-circuit to situations when $\mathcal{T}$ encodes a qubit and is minimally affected by conversion or detection.

\section{Our Platform: Programmable Optical Tweezer Arrays of Laser-Cooled Molecules} 
Our experiment utilizes laser-cooled CaF molecules trapped in a programmable optical tweezer array. A molecular beam of CaF molecules is created using a cryogenic buffer gas cell~\cite{Hutzler2012CBGB}, laser-slowed~\cite{Truppe2017chirp}, and trapped in a magneto-optical trap (MOT)~\cite{Anderegg2017MOT,Truppe2017Mot}. Subsequently, the molecules are further laser-cooled~\cite{Cheuk2018Lambda}, compressed~\cite{Li2023blueMOTCaF}, optically trapped~\cite{Anderegg2018ODT}, and eventually transferred into a programmable 1D array of optical tweezer traps~\cite{Anderegg2019Tweezer,Holland2023bichromatic}. 

We detect CaF molecules via fluorescence imaging; in particular, only molecules in the $X^2\Sigma (v=0, N=1)$ rovibrational manifold can be cooled and imaged, where $v$ ($N$) denote vibrational (rotational) quantum numbers. For fluorescence imaging, light addressing the $X^2\Sigma(v=0, N=1) \rightarrow A^2\Pi_{1/2}(v=0,J=1/2,+)$ transition is applied, and fluorescence emitted on the same transition is collected on a camera. During each image, the molecules explore all 12 hyperfine states in $X^2\Sigma (v=0, N=1)$.

In this paper, no tweezer rearrangement is utilized, but an initial non-destructive image is taken to identify occupied tweezers. This process leaves molecules in the set of detectable states $\mathcal{F}=X^2\Sigma(v=0, N=1)$ with a small probability of molecular loss (leakage into $\left|\emptyset\right \rangle$). All measurements are conditioned on tweezer sites that are initially identified to be occupied. 

\section{New Site-Resolved Detection Scheme with Rapid Resonant Imaging}
\label{sec:errordet}

Compared to detection requirements needed for preparing defect-free arrays, our work here, which concerns enhancing internal state purities and quantum erasure detection, has much more stringent detection requirements. Rather than only needing to preserve the total population in $\mathcal{F}$, we instead need to either preserve the population of a single quantum state or preserve \textit{both population and coherence} within a qubit subspace. These requirements necessitate a new detection scheme, which we describe in this section.

We first discuss the choice of $\mathcal{F}$ and $\mathcal{T}$. Because a measurement projects a qubit, the detection manifold $\mathcal{F}$ must be disjoint from the target state(s) $\mathcal{T}$. In particular, detected population in $\mathcal{F}$ will correspond to an error ($\{e_i\} = \{b_i\}$). In order for a measurement of $\mathcal{F}$ to minimally affect $\mathcal{T}$, a viable approach is to choose internal states for $\mathcal{F}$ and $\mathcal{T}$ such that they have widely separated optical transitions compared to their optical linewidths. This general approach has been proposed and demonstrated in alkaline-earth atomic tweezer arrays~\cite{Wu2022Erasure,Ma2023Erasure,Scholl2023Erasure,Scholl2023HyperEntanglement}, where $\mathcal{F}$ and $\mathcal{T}$ use separate ground and metastable electronic manifolds.

\begin{figure}[t]
	{\includegraphics[width=\columnwidth]{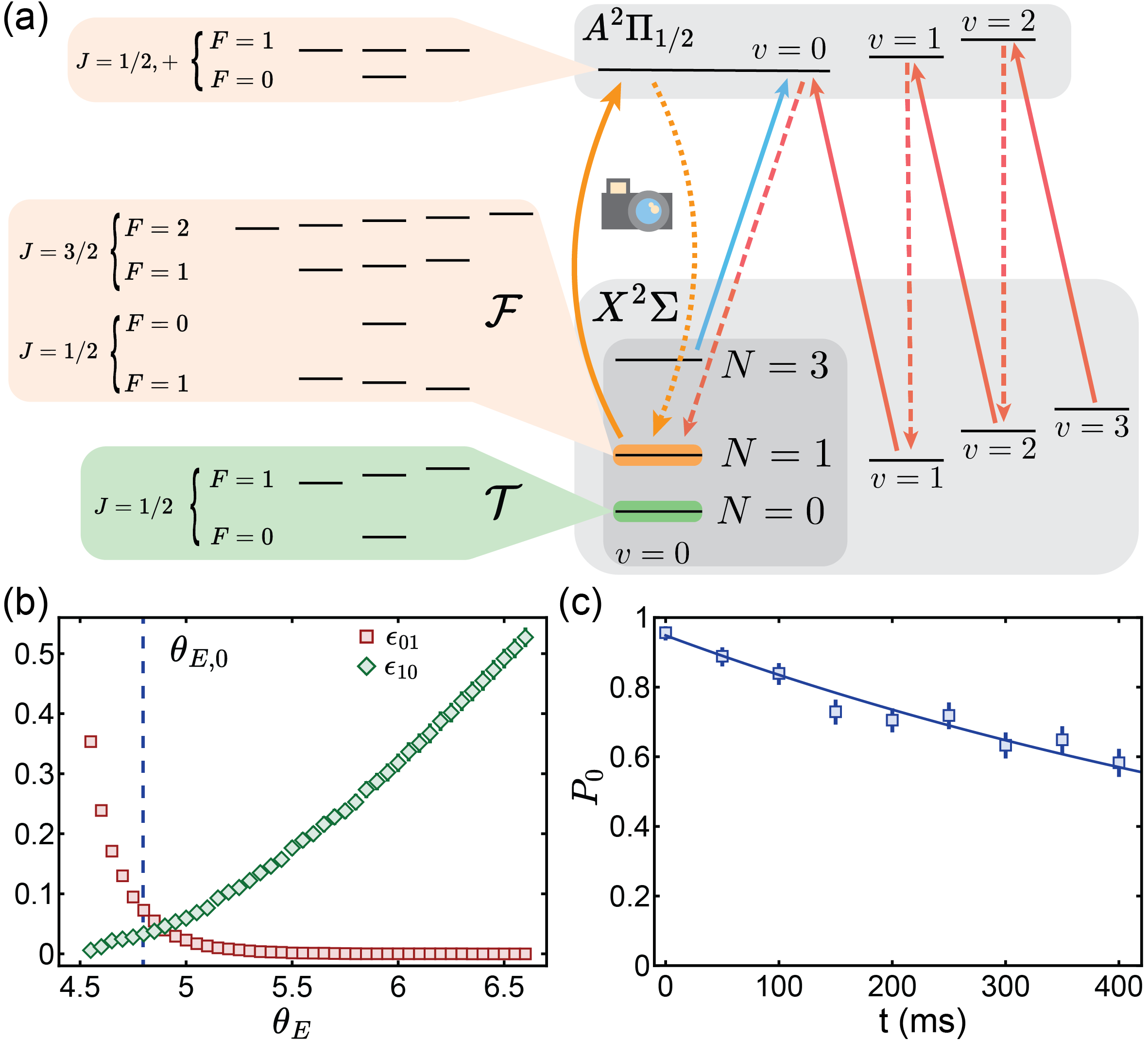}}
	\caption{\label{Fig_1Img} Site-Resolved Detection using Rapid Resonant Imaging. (a) Imaging Scheme. Molecules in the detection manifold $\mathcal{F}=X^2\Sigma(v=0,N=1)$ scatter light resonant with the $X^2\Sigma(v=0,N=1)\rightarrow A^2\Pi_{1/2}(v=0,J=1/2,+)$ transition (orange arrows). Vibrational (red solid arrows) and rotational repumping light (blue solid arrow) is applied. The imaging light is off-resonant for states in the target manifold $\mathcal{T} \subset X^2\Sigma(v=0,N=0)$. (b) Error Identification Performance. The false positive probability $\epsilon_{01}$ (red squares) and false negative probability $\epsilon_{10}$ (green diamonds) are shown versus the classification threshold $\theta_\text{E}$. Vertical dashed line marks $\theta_\text{E}=\theta_\text{E,0}=4.8$, the value used in this work. (c) Population $P_0$ in $\left|{X^2\Sigma(v=0, N=0), J=1/2, F=1,m_F=-1}\right\rangle \in \mathcal{T}$ versus imaging duration $t$. An exponential fit gives a $1/e$ lifetime of $\tau=790(60)\,\text{ms}$, much longer than the imaging duration of $3\,\text{ms}$.}
	\vspace{-0.2in}
\end{figure}

Here, instead of using different electronic manifolds, we make use of two different long-lived rotational manifolds available in a molecule to encode $\mathcal{F}$ and $\mathcal{T}$. Specifically, in our scheme for CaF molecules, we use the set of optical cycling states $X^2\Sigma (v=0, N=1)$ as $\mathcal{F}$, and states from the ground rovibrational manifold $X^2\Sigma (v=0, N=0)$ to form the target subspace $\mathcal{T}$ (Fig.~\ref{Fig_1Img}(a)). The states in $\mathcal{T}$ are separated from the detection manifold states $\mathcal{F}$ by a frequency of $\Delta \approx 2\pi \times 20\,\text{GHz}$, which is much larger than the optical linewidth $\Gamma \approx 2\pi\times 10\,\text{MHz}$ of the relevant imaging transition used to detect $\mathcal{F}$.

We next describe considerations for fluorescence detection of $\mathcal{F}$. We desire to minimize errors for $\mathcal{T}$ when measuring population in $\mathcal{F}$. These can occur due to off-resonant photon scattering of imaging light that affects both populations and coherences in $\mathcal{T}$. A useful figure of merit is the ratio $\eta$ of the off-resonant scattering rate on $\mathcal{T}$ to the fluorescence rate of $\mathcal{F}$. For our choice of $\mathcal{F}$ and $\mathcal{T}$, $\eta$ can be as low as $9\Gamma^2/\Delta^2\sim10^{-6}$ for on-resonant imaging~\cite{Supplement}. Because we scatter $\sim 100-1000$ imaging photons to achieve sufficient signal-to-noise, our scheme has a minimum error of $\sim 10^{-3}$, which predominantly takes the form of leakage into other molecular states. These leakage errors could in principle be corrected with more complex schemes that involve error conversion to $\mathcal{F}$. We note that our on-resonant imaging scheme is different than that previously used for detection in CaF tweezer arrays, where the primary requirement was to preserve molecular population. This was accomplished using light detuned from resonance by $\sim 4\,\Gamma$~\cite{Cheuk2018Lambda,Anderegg2019Tweezer,Holland2023bichromatic}. 

An additional consideration for error detection is crosstalk, where population leaks between $\mathcal{F}$ and $\mathcal{T}$. In our scheme, crosstalk is protected by selection rules. Because the optical transitions used are all E1 transitions, and $\mathcal{F}=X^2\Sigma(v=0,N=1)$ and $\mathcal{T} \subset X^2\Sigma(v=0,N=0)$ are of opposite parity, no population leakage occurs even in the presence of off-resonant scattering from imaging light.

We next describe how we implement on-resonant imaging and optimize its performance. Specifically, to image molecules in $\mathcal{F}$, we apply light resonant with all resolved hyperfine transitions of the $X^2\Sigma(v=0, N=1)-A^2\Pi_{1/2}(v=0, J=1/2,+)$ line. Laser light addressing the $X^2\Sigma(v=1,2,3, N=1) \rightarrow A^2\Pi_{1/2}(v'=v-1, J=1/2,+)$ transitions acting as vibrational repumpers is also applied. We additionally apply light addressing the $X^2\Sigma(v=0, N=3) \rightarrow B^2\Sigma(v'=0, N=2)$ transition, which acts as a rotational repumper~\cite{Holland2023bichromatic}. These beams propagate perpendicular to the imaging axis, and cause no observable background for typical imaging durations. 

To optimize imaging for the purpose of identifying errors, we prioritize minimizing the false negative probability ($\epsilon_{10}$), since it is the probability that errors remain undetected. We note that for tweezer preparation, the false positive probability ($\epsilon_{01}$) only affects the data rate, since any site flagged as an error could be discarded during subsequent rearrangement or post-selection. We therefore seek to optimize $\epsilon_{10}$ with respect to the imaging light intensity ($X^2\Sigma(v=0, N=1)-A^2\Pi_{1/2}(v=0, J=1/2,+)$), the imaging duration, and the tweezer depth. Specifically, we minimize $\epsilon_{10}$ at a fixed threshold of $\theta_\text{E}=\theta_\text{E,0}=4.8$, and achieve a false positive probability ($\epsilon_{10}(\theta_\text{E,0})=0.033(4)$) at an imaging intensity of $I=4.5\,\text{mW/cm}^2$, imaging duration of $3\,\text{ms}$, and tweezer depth of $k_B\times930(20)\,\mu\text{K}$ (Fig.~\ref{Fig_1Img}(b)). 
In principle, for detectable errors, these values allow reduction of intrinsic error rates by up to a factor of $\sim 30$. Henceforth, $\theta_\text{E}$ is fixed to $\theta_\text{E,0}$ unless noted. Because the background is low enough to be largely independent of the imaging duration, $\epsilon_{01}$, which determines the data rate, is independent of imaging parameters and only depends on camera noise.

To verify that our site-resolved detection scheme minimally affects population in $\mathcal{T}$, which is composed of states from $X^2\Sigma(v=0,N=0)$, we prepare molecules in $|X^2\Sigma(v=0,N=0), J=1/2, F=1, m_F=-1\rangle$ and measure the population loss per image (Fig.~\ref{Fig_1Img}(c)). We find a loss rate of $3.8(3)\times10^{-3}$ per image, confirming that our scheme minimally affects the population in $X^2\Sigma(v=0,N=0)$.

\begin{figure}[t]
	{\includegraphics[width=\columnwidth]{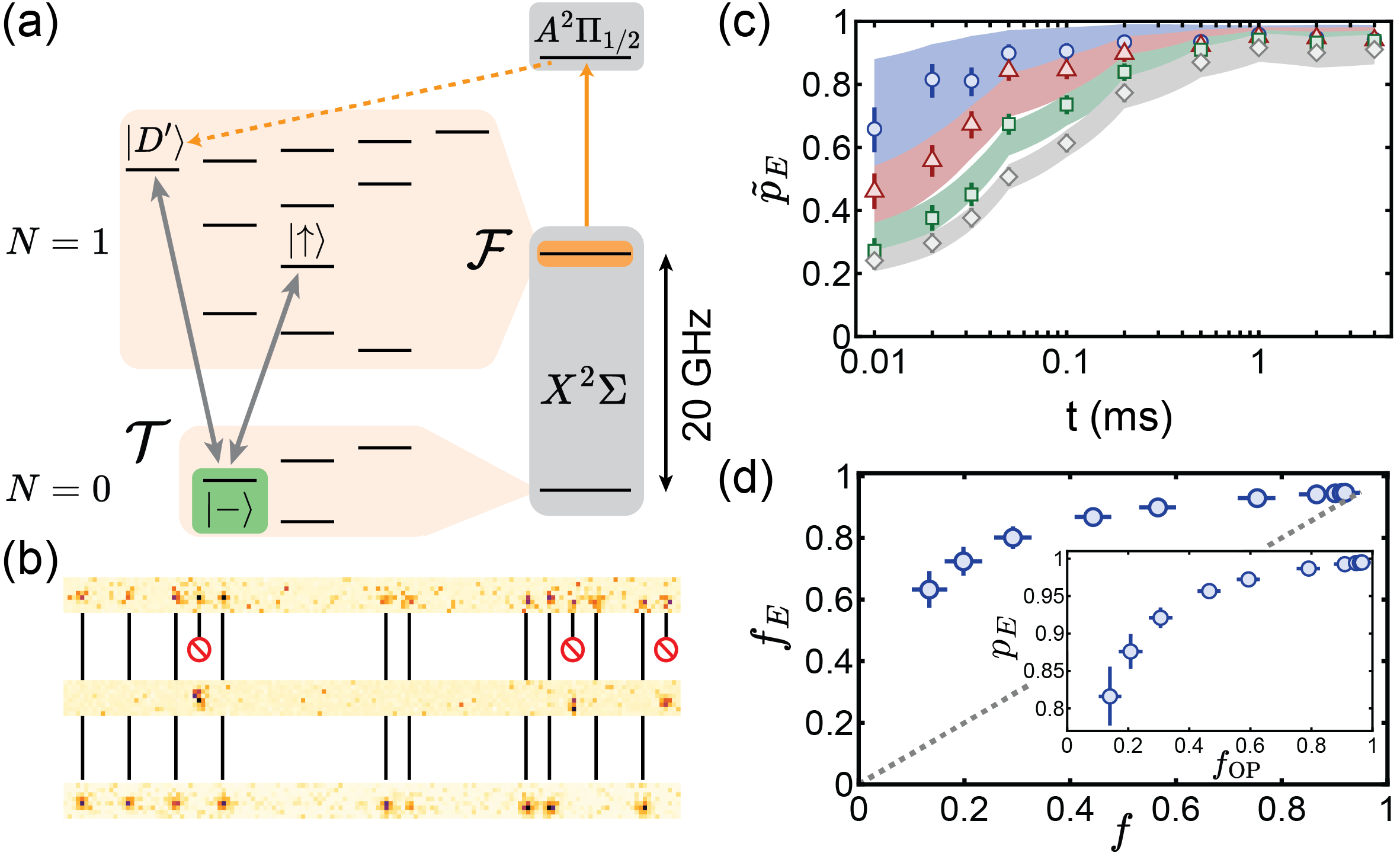}}
	\caption{\label{Fig_1SP} Robust and Enhanced Tweezer Preparation Using Site-Resolved Error Detection. (a) Our enhanced internal state preparation procedure involves: (i) optical pumping to $|D'\rangle$ (orange arrow), (ii) population transfer to $\mathcal{T}=|-\rangle$ via a microwave pulse (grey arrow), and (iii) error detection of left-over molecules in $\mathcal{F}$. 
	(b) Verifying Tweezer Preparation. A non-destructive image (top) identifies initially loaded sites. The detection-enhanced internal state preparation procedure is applied, resulting in an error image (middle) that flags sites with internal state errors. For verification, we state-selectively transfer molecules from $\mathcal{T}=\left|-\right\rangle$ to $\left|\uparrow\right\rangle$ for a final destructive measurement (bottom). (c) The detection-conditioned probability $\tilde{p}_E$ of a molecule appearing in the final image versus optical pumping duration $t$, which is used to control the bare optical pumping fidelity $f_{\text{OP}}$. $\tilde{p}_E$ for various error-detection image classification thresholds $\theta_E=(4.6,5.6,6.6,\infty)$ (blue circles, red triangles, green squares, gray diamonds) are shown. Colored bands show corresponding theoretical predictions. (d) Error-excised tweezer fidelity $f_E$ (probability of being occupied and in the target internal state) versus bare tweezer fidelity $f$. Inset shows the error-excised internal state purity $p_E$ versus the optical pumping fidelity $f_{\text{OP}}$.}
	\vspace{-0.2in}
\end{figure}

\section{Robust and Enhanced Internal State Preparation Using Site-Resolved Detection of Errors}
We next make use of our site-resolved error detection scheme to achieve robust and enhanced tweezer preparation fidelities. In previous work, the fidelity of having a tweezer loaded with a molecule in a target internal state was reported to be $\approx 80\%$, limited both by optical pumping fidelities and imperfect microwave transfers~\cite{Holland2023Entanglement,Bao2023Entanglement}. 

Here, we present a scheme that allows internal state errors to appear as detectable errors, which can be subsequently corrected mid-sequence. Our scheme works as follows (Fig.~\ref{Fig_1SP}(a)). Initially, molecules are distributed over all 12 hyperfine states in $X^2\Sigma (v=0, N=1)$, the laser-coolable manifold, which is also the detection manifold $\mathcal{F}$. We optically pump the molecules into a single hyperfine state $|D'\rangle=|X ^2\Sigma(v=0, N=1), J=3/2, F=2, m_F=-2\rangle  \in \mathcal{F}$. Subsequently, we ``shelve'' them into the ground rotational manifold $X^2\Sigma (v=0, N=0)$. Specifically, we use a microwave pulse with a frequency of $\approx 20.5\,\text{GHz}$ to state-selectively transfer $|D'\rangle$ to a single internal state $|-\rangle=|X^2\Sigma(v=0,N=0),J=1/2,F=1,m_F=-1\rangle$, which is our target state ($\mathcal{T}=\{|-\rangle \}$). All internal state preparation errors leave the molecule in $\mathcal{F}$, that is, $\mathcal{E}=\mathcal{F}$. Specifically, optical pumping errors leave a molecule in $(\mathcal{F}-\left|D'\right\rangle)\subset \mathcal{F}$, and imperfect microwave transfer leaves a molecule in $\left|D'\right\rangle \in \mathcal{F}$. Internal state preparation errors thus appear directly as detectable errors. They can then be identified by the site-resolved detection scheme described in Section~\ref{sec:errordet}, which minimally affects population in $\mathcal{T}=\{|-\rangle\}$ on error-free sites.

We next implement our scheme and examine its performance. Following internal state preparation into $\mathcal{T}$, we perform one error detection image. Subsequently, we transfer molecules from the target state $\left|-\right\rangle$ to $|{\uparrow}\rangle=\left|X^2\Sigma(v=0, N=1), J=1/2, F=0, m_F=0)\right\rangle$, which is part of the detection manifold $\mathcal{F}$ (Fig.~\ref{Fig_1SP}(b)). Finally, we destructively measure the total population in $\mathcal{F}$. 

To evaluate the robustness of our scheme, we intentionally vary the amount of optical pumping errors by varying the optical pumping duration $t$. For each optical pumping duration, which serves as a proxy for the optical pumping fidelity, we extract $\tilde{p}_E$, the probability that we detect a molecule in the final image conditioned upon no identified error. In (Fig.~\ref{Fig_1SP}(c)), we show $\tilde{p}_E$ for various error detection thresholds $\theta_E$. As expected, identifying errors more aggressively by lowering $\theta_E$ (smaller false-negative probability $\epsilon_{10}$ and larger false-positive probability $\epsilon_{01}$) improves $\tilde{p}_E$. 

To show the gains of our scheme directly, we compare the internal state purity conditioned upon no errors, $p_\text{E}$, with the optical pumping fidelity $f_{\text{OP}}$. We determine the experimentally achieved $f_{\text{OP}}$ for each pumping duration $t$ using $\tilde{p}_E$ and the error imaging infidelities $\epsilon_{01}$ and $\epsilon_{10}$~\cite{Supplement}. As shown in Fig.~\ref{Fig_1SP}(d, inset), error excision allows significantly improved internal state purity, especially at low optical pumping fidelities. Notably, we achieve an internal state purity of $p_E=0.995(1)$ when the achieved bare optical pumping fidelity is $f_{\text{OP}}\approx 0.96$. Notably, $p_E$ remains high even when $f_{OP}$ is degraded. For example, at an optical pumping fidelity of $f_{\text{OP}}=0.46$, the error-excised internal state purity remains at $p_E\approx 0.96$. Finally, we extract the error-excised tweezer preparation fidelity $f_E$, defined as the probability that a tweezer site is occupied with a molecule in the target internal state when conditioned upon an absence of a detected error. This is the practically useful metric for preparing low-defect arrays using tweezer rearrangement. In Fig.~\ref{Fig_1SP}(d), we show $f_E$ versus the bare tweezer preparation fidelity $f$, which takes into account occupation errors. As expected, we find that error excision provides significant improvement and robustness. Notably, we achieve a tweezer preparation fidelity of $f_E=0.952(3)$, with an estimated reduction in data rate of only $7.2(2)\%$. This significantly improves upon previous reported values ($f\approx 80\%$)~\cite{Holland2023Entanglement,Bao2023Entanglement,Lu2024RSC,Bao2023RSC} and importantly, our method is robust to day-to-day fluctuations in $f_{\text{OP}}$.

We note although all error-conditioned measurements in this work are done via post-selection, where we excise sites that are error-flagged, our measurements indicate that our detection scheme minimally affects correctly prepared molecules. Thus, in combination with the ability to perform low-loss rearrangement (previously demonstrated to have losses $\sim 0.1\%$~\cite{Holland2023Entanglement}), our scheme can be used with another round of tweezer rearrangement to achieve similar final fidelities. This will allow the creation of arrays with low defect rates both in spatial configuration and internal state purity, which is crucial to quantum computation and simulation applications with large-scale molecular tweezer arrays. In particular, the ability to create arrays with few-percent level errors opens the door to simulating interacting quantum spin systems with low defect rates. Specifically, these low defect rates would allow practical post-selection for perfect arrays (possible with full state-resolved detection) for up to 10 sites, with data rates reduced by only $\sim 40\%$.

\section{A Quantum Erasure Detection Scheme For CaF Molecules}

\begin{figure}[t]
	{\includegraphics[width=\columnwidth]{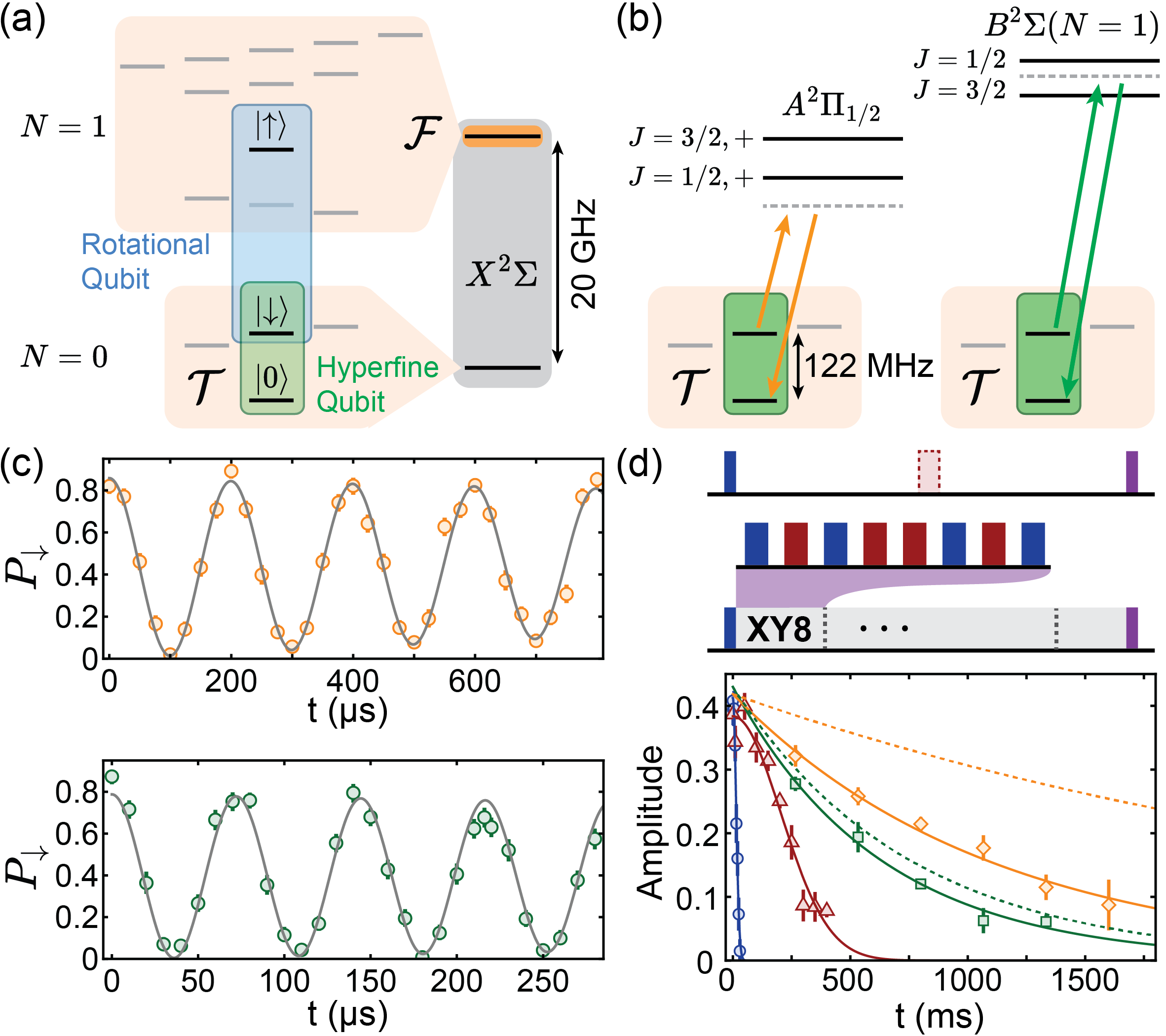}}
	\caption{\label{Fig_2} Coherent Control and Coherence Times of New Hyperfine Qubit. (a) Previous rotational qubit encoding $\left|{\downarrow}\right\rangle$-$\left|{\uparrow}\right\rangle$ (boxed in blue) is not compatible with our error detection scheme, since $\left|\uparrow\right\rangle$ is in the detection manifold $\mathcal{F}$. Our new qubit encoding $\left|{\downarrow}\right\rangle$-$\left|0\right\rangle$ uses two hyperfine states in $X^2\Sigma(v=0,N=0)$ that are outside of $\mathcal{F}$ (boxed in green). (b) Coherent manipulations of $|{\downarrow}\rangle$-$|0\rangle$ is performed using a two-photon Raman processes. Our two schemes rely on light near-detuned to the $X$-$A$ and $X$-$B$ optical transitions. (c) Molecules are prepared in $\left|\downarrow\right\rangle$ and driven using Raman light for a variable duration $t$. The top (bottom) panels show Rabi oscillations of the $\left|\downarrow\right\rangle$ population when driven by $X$-$A$ ($X$-$B$) Raman light. Solid lines show fits to damped sinusoids. (d) Hyperfine qubit coherence versus hold time is measured via a Ramsey sequence (top panel). Data points show measured Ramsey fringe amplitudes versus hold time $t$, with solid lines showing fits to exponential decay curves. Data without (with) a spin-echo pulse are shown by blue circles (red triangles); data with XY8 dynamical decoupling pulses using $X$-$B$($X$-$A$) light are shown by green squares (orange diamonds). The fitted $1/e$ lifetimes are $19.5(7)\,\text{ms}$, $288(15)\,\text{ms}$, $630(50)\,\text{ms}$, and $1100(70)\,\text{ms}$, respectively. The dashed green (orange) line shows the estimated population decay due to off-resonant scattering for the $X$-$B$ ($X$-$A$) Raman configuration~\cite{Supplement}.}
	\vspace{-0.2in}
\end{figure}

In the second part of our work, we demonstrate erasure detection for the first time in molecules. Compared to enhancing tweezer preparation using site-resolved detection, where only the total population in $\mathcal{F}$ needs to be preserved, erasure detection is more challenging since both population and coherence in error-free qubits should be minimally affected.

\subsection{A New Hyperfine Qubit Encoding: Coherent Control and Qubit Coherence}
To implement quantum erasure detection, we first seek a qubit encoding that is compatible with our site-resolved error detection scheme described in Section~\ref{sec:errordet}. In previous work with CaF molecules, qubits were encoded using two states from neighboring rotational manifolds, namely, $|{\uparrow}\rangle=|X^2\Sigma(v=0,N=1), J=1/2, F=0, m_F=0\rangle$ and $|{\downarrow}\rangle=|X^2\Sigma(v=0,N=0), J=1/2, F=1, m_F=0\rangle$~\cite{burchesky2021rotcoh,Holland2023Entanglement,Bao2023Entanglement}. This rotational encoding was used to demonstrate a 2-qubit iSWAP gate, but is incompatible with our detection scheme because $|{\uparrow}\rangle$ is part of the detection manifold $\mathcal{F}$. The detection of $\mathcal{F}$, which is a projective measurement of its population, will destroy qubit coherence between $|{\uparrow}\rangle$ and $|{\downarrow}\rangle$. 

Here, we use a new qubit encoding that makes use of two hyperfine states within the ground rovibrational manifold $X^2\Sigma(v=0,N=0)$, which are dark to the imaging light used for site-resolved error detection. Specifically, the hyperfine encoding uses the states $|{\downarrow}\rangle=|X^2\Sigma(v=0,N=0), J=1/2,F=1,m_F=0\rangle$ and $|0\rangle=|X^2\Sigma(v=0,N=0), J=1/2,F=0,m_F=0 \rangle$ (Fig.~\ref{Fig_2}(a)). These states are predicted to have long coherence times because of the absence of tensor ac Stark shifts and their low magnetic moments. In the framework of Section~\ref{sec:framework}, the detection manifold $\mathcal{F}$ remains the same ($\mathcal{F}=X^2\Sigma(v=0,N=1)$), while the target subspace $\mathcal{T}= \{|{\downarrow}\rangle,|{0}\rangle\}$ now consists of the two hyperfine qubit states.

We first demonstrate coherent control over this hyperfine qubit, and that it has long coherence times. The two qubit states are connected by an M1 transition at $\sim 122\,\text{MHz}$. We coherently manipulate the qubit with two-photon Raman transitions rather than directly driving the M1 transition. Specifically, we use Raman light near-detuned to either the $X-A$ or $X-B$ optical transitions (Fig.~\ref{Fig_2}(b)). The two frequencies in the Raman light co-propagate and drive $\sigma^{+}$ transitions. To demonstrate coherent control, we prepare molecules in $|{\downarrow}\rangle$, and apply Raman light for a variable duration. As shown in Fig.~\ref{Fig_2}(c), we observe Rabi oscillations with Rabi frequencies as high as $\Omega_R\approx2\pi\times10\,\text{kHz}$ for both Raman schemes.

We next measure the qubit coherence time using a Ramsey spectroscopy sequence consisting of two Raman $\pi/2$ pulses separated by a variable duration. In order to minimize the effect of inhomogeneous light shifts among the tweezer sites, we operate at a tweezer depth of $U_{\text{hf}}=k_B\times 39(1)\,\mu\text{K}$. By measuring the decay rate of the amplitude of the Ramsey fringe, we find a bare coherence time of $T_2^*=19.5(7)\,\text{ms}$, which can be increased to $T_2=288(15)\,\text{ms}$ with a single spin-echo. The coherence time can be further extended by applying dynamical decoupling. Specifically, using a XY8 pulse sequence (pulse separation of 33.3\,\text{ms}), the coherence times increase to $630(50)\,\text{ms}$ and $1100(70)\,\text{ms}$ for the $X$-$B$ Raman scheme and $X$-$A$ Raman scheme, respectively (Fig.~\ref{Fig_2}(d)).

In addition to long coherence times, we note that our hyperfine qubit encoding is also compatible with the previously used rotational qubit encoding~\cite{burchesky2021rotcoh,Holland2023Entanglement,Bao2023Entanglement}. We have verified that the three states involved can remain simultaneously coherent, and that we can coherently transfer information encoded in a rotational qubit ($|{\uparrow}\rangle, |{\downarrow}\rangle$) to a hyperfine encoding ($|{\downarrow}\rangle, |{0}\rangle$)~\cite{Supplement}.

\begin{figure}[t]
	\includegraphics[width=\columnwidth]{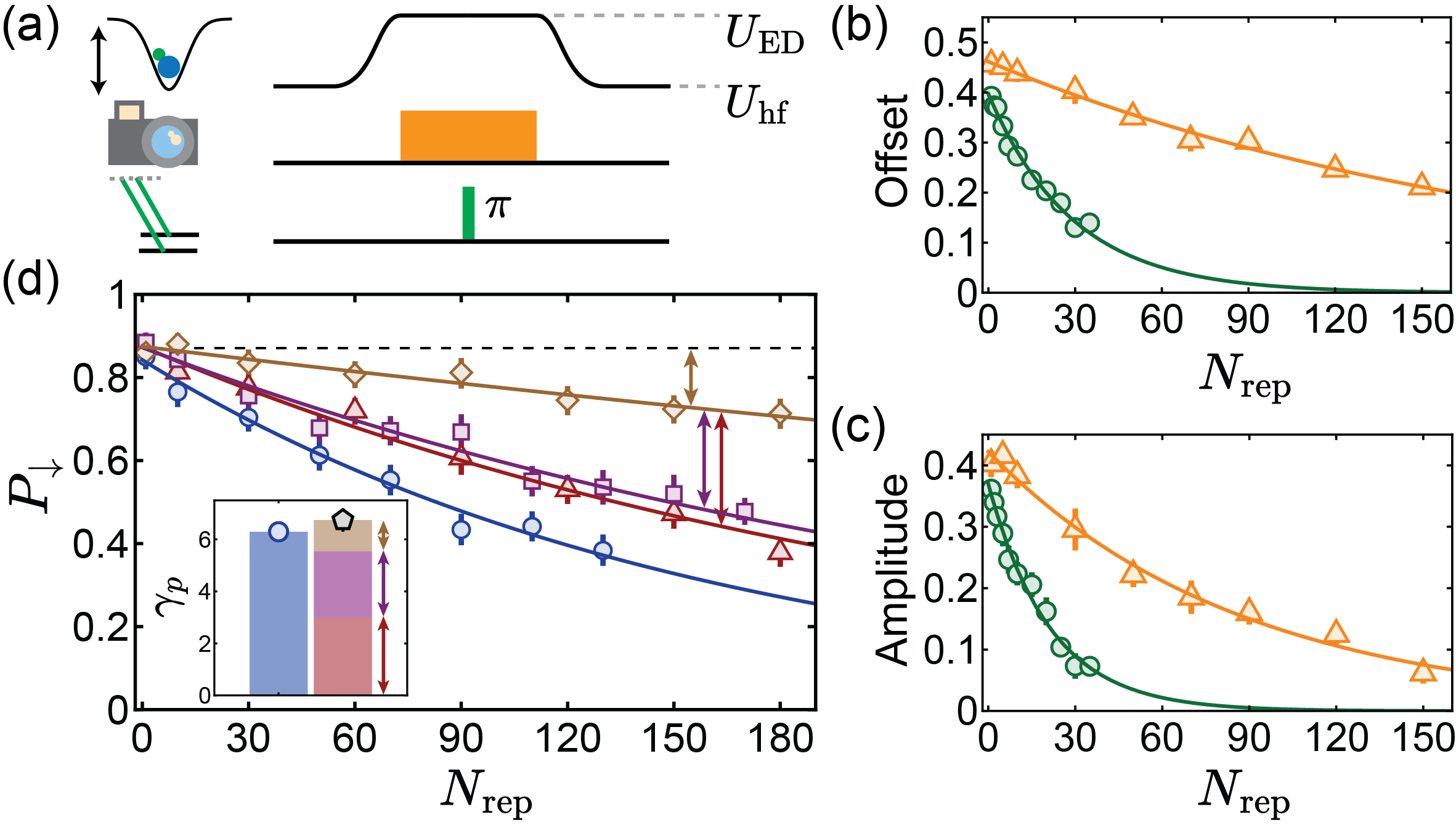}
	\caption{\label{Fig_3} Mid-Circuit Quantum Erasure Detection. (a) Our composite erasure detection scheme comprises a tweezer depth increase, an error detection image, and a mid-detection $X$-$B$ Raman $\pi$-pulse. (b) Ramsey fringe offset versus the number of composite erasure detections $N_{\text{rep}}$. (c) Ramsey fringe amplitude versus $N_{\text{rep}}$. For (b,c), green circles (orange triangles) are data obtained using the $X$-$B$ ($X$-$A$) Raman scheme.(d) Population loss measurements from combinations of tweezer ramp (T), error detection light (E), and $v=1$ repump (V). The plot shows population $P_{\uparrow}$ versus $N_{\text{rep}}$ cycles of TEV (blue circles), EV (red triangles), T (purple squares).  Brown diamonds are measured at a shallow tweezer depth without TEV. The brown, purple, and red arrows indicate contributions from background vacuum and blackbody loss, tweezer ramp, and erasure detection. Inset: The sum of these measured loss rate contributions (colored bars summing to gray pentagon) is consistent with the total loss (blue bar and circle).}
	\vspace{-0.2in}
\end{figure}

\subsection{Composite Erasure Detection Sequence to Preserve Qubit Coherence}
Having established long coherence times and coherent control of the hyperfine qubit, we next investigate whether qubit coherence is preserved after an error detection image. Compared to the tweezer depth $U_{\text{hf}}$ used to obtain long hyperfine qubit coherence times, we find that a much higher tweezer depth of $U_{\text{ED}}=k_B\times930(20)\,\mu\text{K}$ is needed to obtain high error detection fidelities. 

Therefore, for erasure detection, the tweezer trap needs to be ramped up by a factor of $\sim30$ from $U_{\text{hf}}$ to $U_{\text{ED}}$. At the deeper depth, the qubit frequency shifts by $2\pi\times0.40(6)\,\text{kHz}$, which on average leads to additional phase accumulation in the hyperfine qubit. This phase accumulation is inconsequential if the tweezers had uniform depth. However, our tweezer array has depth inhomogeneity of $\sim10\%$, giving rise to an effective phase variation of $\approx1\,\text{rad}$ at $U_{\text{ED}}$ over the imaging duration. To counteract this inhomogeneous dephasing, we implement a ``composite erasure detection" scheme consisting of four steps (Fig.~\ref{Fig_3}(a)): 1) the tweezer depth is ramped from $U_{\text{hf}}$ to $U_{\text{ED}}$, 2) local errors are detected using rapid resonant imaging, 3) a $\pi$-pulse is applied to the qubit halfway through the image, and  4) the tweezer depth is returned to $U_{\text{hf}}$. The $\pi$-pulse echos out the effects of inhomogeneous light shifts due to tweezer depth variations, along with Stark shifts arising from the imaging light.

To determine how well the qubit population and coherence are preserved during composite erasure detection, we repeatedly apply composite erasure detection between two $\pi$/2 pulses in a Ramsey sequence. We find that a Ramsey fringe is observable even after multiple images. The offset of the Ramsey fringe is proportional to the total population, and the amplitude is proportional to both the total population and coherence of the qubit states. Per image, we determine a fractional population reduction of $\varepsilon_{p}=3.3(2)\times10^{-2}$ (Fig.~\ref{Fig_3}(b)) and a fractional amplitude reduction of $\varepsilon_{c}=4.6(3)\times10^{-2}$ (Fig.~\ref{Fig_3}(c)). These measurements indicate that our composite erasure detection sequence affects the hyperfine qubit population and coherence only at the few percent level. To our knowledge, this is the first demonstration of erasure detection in molecules. 

We note that for the purposes of composite erasure detection, we only use $X-B$ Raman light, which has much shorter population lifetimes and coherence times compared to our $X-A$ Raman scheme due to higher off-resonant scattering. This is because our $X-A$ Raman light (due to constraints on laser availability) is resonant with the detection manifold $\mathcal{F}=X^2\Sigma(v=0,N=1)$. It resonantly heats a molecule in $\mathcal{F}$ out of a tweezer, and is therefore not compatible with erasure detection. This limitation can easily be overcome with a $X-A$ Raman light source at a more optimal detuning. As we describe in detail in the next section, we anticipate that errors at the $10^{-3}$ level are achievable.

\subsection{Technical and Fundamental Limits of Our Erasure Detection Scheme}
\label{sec:techAndFundLimits}
To understand the origin of the $\sim 10^{-2}$ level errors per erasure image, we independently measure the contributions of each step in our composite erasure detection sequence. First, we quantify the effect of
off-resonant photon scattering on the qubit states caused by the spin-echo $\pi$-pulse. We perform a similar Ramsey measurement with only $\pi$ pulses applied (without tweezer ramps and imaging light). This gives a population error per image of $\varepsilon_{p,R}=2.7(2)\times 10^{-2}$ and a coherence error per image of $\varepsilon_{c,R}=4.4(4)\times10^{-2}$. Comparing these with the total composite erasure errors $\varepsilon_p$ and $\varepsilon_c$, we conclude that our present scheme is limited by off-resonant scattering of Raman light by the qubit states in $\mathcal{T}$. 

Off-resonant scattering can be significantly improved by a more optimal Raman scheme. In particular, we expect that coherence and population errors could be reduced to the $10^{-3}$ level. To quantify the off-resonant scattering rate, we define a Raman $\pi$-pulse quality factor $Q_\pi=\Omega_R/\pi\Gamma_{sc}$, where $\Omega_R$ is the two-photon Rabi frequency, and $\Gamma_{sc}$ is the photon scattering rate. Because off-resonant scattering predominantly leads to leakage from the qubit subspace, $\Gamma_{sc}$ can be directly inferred from population loss rates. $Q_\pi$ is therefore directly measurable. In our present Raman scheme ($X$-$B$), the light is detuned halfway between the $X^2\Sigma(N=0) \rightarrow B^2\Sigma(N=1,J=1/2)$ and $X^2\Sigma(N=0) \rightarrow B^2\Sigma(N=1,J=3/2)$ transitions. This results in a single-photon detuning of $\Delta_{XB}\sim2\pi\times500\,\text{MHz}$ for optical transitions of the qubit states in $\mathcal{T}$. $Q_\pi$ for this scheme is measured to be 37(3).

To partially investigate a better Raman scheme, we performed similar Ramsey measurements using an $X$-$A$ Raman scheme, where the Raman light is detuned by $\Delta_{XA}\sim2\pi\times20\,\text{GHz}$ from the optical transitions of the hyperfine qubit. This light cannot be used for composite erasure detection because it is resonant with the detection manifold $\mathcal{F}$. Nevertheless, we can quantify its effect on the hyperfine qubit. A Ramsey measurement reveals significantly lower population errors and coherence errors per $\pi$-pulse. Specifically, $\varepsilon_{p,R}=3.4(3)\times10^{-3}$ and $\varepsilon_{c,R}=10(1) \times 10^{-3}$. When this scheme is employed in a composite erasure detection sequence (with tweezer ramps but no erasure image due to technical limitations), we obtain overall population and coherence errors of $\varepsilon_p=5.2(2)\times10^{-3}$ and $\varepsilon_c=11(1)\times10^{-3}$ per image, respectively (Fig.~\ref{Fig_3}(b,c)). These improvements agree with predictions due to off-resonant scattering~\cite{Supplement}.

Our current $X$-$A$ Raman scheme could thus be improved significantly and made compatible with composite erasure detection simply by using a different detuning. In particular, by detuning halfway between the $X^2\Sigma(v=0) \rightarrow A^2\Pi_{1/2}(v=0)$ and the $X^2\Sigma(v=0) \rightarrow A^2\Pi_{3/2}(v=0)$ transitions ($\sim1.1\,\text{THz}$ from each), the Raman light will no longer be resonant with the detection manifold $\mathcal{F}$. The number of off-resonantly scattered photons for a Raman $\pi$-pulse would then decrease by a factor of 30 compared to our current $X$-$A$ scheme. Conservatively, we estimate that population and coherence errors due to Raman light at an optimal detuning will be suppressed to $\varepsilon_{p,R} \approx 1\times 10^{-4}$ and $\varepsilon_{c,R} \approx 6\times 10^{-3}$. 

When the off-resonant Raman scattering errors are reduced to these levels, we must examine the remaining error sources. These include the detection light, the tweezer depth ramps, and loss to higher vibrational states due to blackbody radiation. We measure their individual contributions by comparing population loss from $|{\downarrow}\rangle$ with and without various combinations of detection light, tweezer ramps, and $v=1$ vibrational repumping light (Fig.~\ref{Fig_3}(d)). From these measurements, we isolate a population loss per image of $3.0(3)\times10^{-3}$ due to detection light, and $2.6(3)\times10^{-3}$ due to the tweezer ramps. The background loss in the absence of detection light and tweezer ramps is $1.2(1)\times10^{-3}$, consistent with the predicted excitation rate to higher vibrational states due to blackbody radiation at room temperature. Therefore, by using optimally detuned Raman light, we estimate that total population (coherence) errors of $7\times10^{-3}$ ($1.4\times10^{-2}$) per image can be achieved. 

The population loss rates can be further improved. First, improving the photon collection efficiency of the imaging system allows shorter image durations and therefore lower population loss. Second, the population loss appears primarily as internal state leakage errors into the $X^2\Sigma(v=0, N=2)$ manifold and the $\left|X^2\Sigma(v=0, N=0), J=1/2, F=1, m_F=\pm 1\right\rangle$ states. These errors can be erasure-converted to $\mathcal{F}$ via additional microwave and optical pulses. Thus, we estimate that the population error due to detection light can be suppressed to well below $10^{-3}$. Furthermore, blackbody induced errors can be erasure-converted, as we will show, or eliminated in a cryogenic setup. Depending on the population loss mechanism of the tweezer ramps, which we did not explore, the corresponding population errors could potentially be converted and corrected. Therefore, population errors at the low $10^{-3}$ level could potentially be achievable.

\begin{figure}[t]
	\includegraphics[width=\columnwidth]{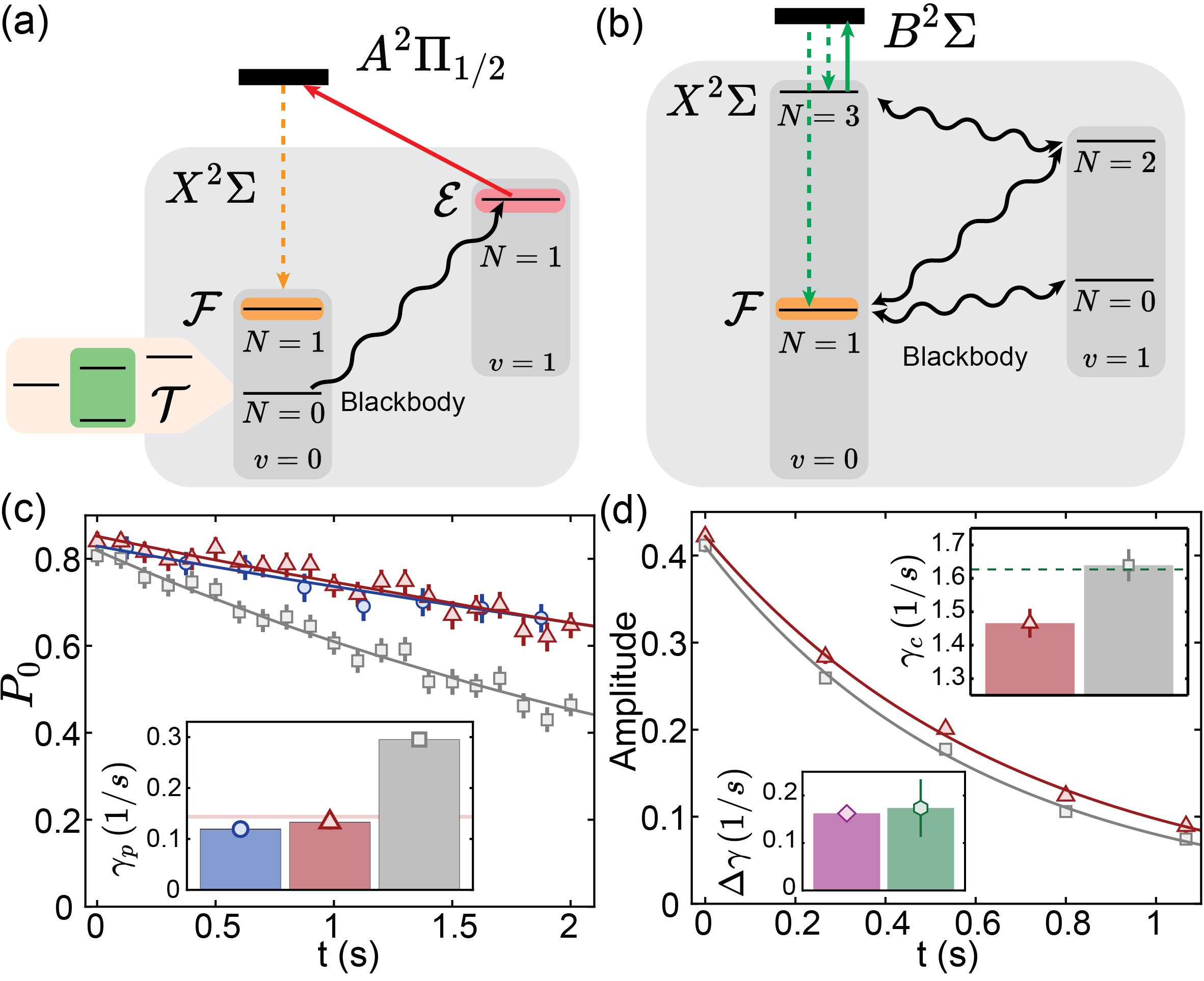}
	\caption{\label{Fig_4} Erasure Conversion of Blackbody Errors. (a) Erasure Conversion Scheme. Blackbody radiation (black wavy arrow) predominantly excites molecules from $\mathcal{T}$ to $\mathcal{E}$. We erasure convert to $\mathcal{F}$ using $v=1$ repumping light (red arrow). (b) Robustness of $\mathcal{F}$ to blackbody leakage errors. Blackbody leakage errors (black wavy) primarily populate $X^2\Sigma(v=0,N=3)$ states, which are converted back to $\mathcal{F}$ by the rotational repumping light present during erasure detection (solid green). (c) Hyperfine qubit population $P_0$ versus hold time $t$. Data without erasure excision (gray squares), with a single final error detection image (red triangles), and with excision using five mid-sequence images (blue circles) are shown; solid lines show fits to exponential decay curves. Inset: fitted $1/e$ loss rates $\gamma_p$ are shown, with red band ($\pm1\sigma$) indicating prediction from separately measured rates~\cite{Supplement}. No dynamical decoupling Raman pulses are applied. (d) Ramsey fringe amplitude of the hyperfine qubit versus hold time with (red triangles) and without (gray squares) error excision. Dynamical decoupling with Raman pulses is applied during the hold time. Upper Inset: Amplitude loss rate $\gamma_c$ without error excision (gray square) is consistent with the expected combined loss rate from the $X$-$B$ Raman beams and the $N=0$ population loss (green dashed line). Lower inset shows the loss rate improvement $\gamma_c$ when excising errors. The improvement for the amplitude loss rate (green hexagon) is consistent with the improvement in population loss (purple diamond).}
	\vspace{-0.2in}
\end{figure}

\section{Quantum Erasure Conversion and Detection of Blackbody Errors}

We next demonstrate that leakage errors induced by blackbody radiation can be converted and detected as erasures. In CaF, blackbody radiation at room-temperature predominantly drives $\Delta v=1$ vibrational-changing E1 transitions, and leads to population leakage from the vibrational ground manifold $X^2\Sigma(v=0)$ to the first excited vibrational manifold $X^2\Sigma(v=1)$ at a rate of $\Gamma_{01}\approx0.3\,\text{s}^{-1}$~\cite{hou2018VibDip}. Consequently, the hyperfine qubit states $\mathcal{T}= \{|{\downarrow}\rangle, |0\rangle\} \subset X^2\Sigma(v=0,N=0)$ predominantly leak to $\mathcal{E}=X^2\Sigma(v=1,N=1)$. As we will show, these errors can be erasure converted by using $v=1$ vibrational repumping light, which excites molecules from $\mathcal{E}$ to $A^2\Pi_{1/2}(v=0,J=1/2,+)$. The excited molecules decay back to $\mathcal{F}$ with $98\%$ probability (Fig.~\ref{Fig_4}(a)).

We note that $\mathcal{E}$ spontaneously decays back to $X^2\Sigma(v=0)$ at a rate $\Gamma_{10}\approx5.3\,\text{s}^{-1}$, which is much quicker than the blackbody excitation rate $\Gamma_{01}$. If such a decay occurs, conversion via $v=1$ vibrational repumping light is no longer possible. Therefore, in our scheme, one must erasure convert $\mathcal{E}$ to $\mathcal{F}$ at a rate much faster than $\Gamma_{10}$. It is not necessary, however, to detect $\mathcal{F}$ faster than $\Gamma_{10}$.

We next discuss the effect of blackbody radiation on the detection manifold $\mathcal{F}$. Absorption of a blackbody photon brings a molecule from $\mathcal{F}$ to $X^2\Sigma(v=1,N=0,2)$. With a $1/e$ decay time of $1/\Gamma_{10} \approx 190\,\text{ms}$, these states spontaneously decay back to $\mathcal{F}$ and $X^2\Sigma(v=0,N=3)$. Molecules in the latter manifold can be pumped back to $\mathcal{F}$ via rotational repumping light, which is already present during fluorescence imaging. Therefore, even in the presence of blackbody radiation, there is minimal undetectable leakage from $\mathcal{F}$ (Fig.~\ref{Fig_4}(b)).

Experimentally, we first confirm that population in $\mathcal{F}$ is robust to blackbody induced leakage errors. As expected, when a rotational repump is applied during detection, we find that the population lifetime ($\tau=13.5(5)\,\text{s}$) exceeds the blackbody-limited $X^2\Sigma(v=0)$ lifetime by a factor of $\approx4$. We attribute residual loss to background gas collisions and double blackbody excitation events into $N\geq5$, which are not rotationally repumped during detection~\cite{Supplement}. 

Next, we investigate detection of population leakage errors by preparing molecules in each of the two hyperfine levels in $\mathcal{T}$ and comparing population loss rates with and without excising errors. Specifically, we hold the molecules for a variable duration, and perform mid-sequence conversion by applying $v=1$ repumping light every $50\,\text{ms}$. Subsequently, we perform site-resolved error detection, and destructively detect population remaining in the initially prepared hyperfine state. Without error excision, we measure a population loss rate of $\gamma_p=0.30(1)\,\text{s}^{-1}$, consistent with the predicted blackbody excitation rate. With error excision, the loss rate decreases to $\gamma_{p,E}=0.13(1)\,\text{s}^{-1}$ (Fig.~\ref{Fig_4}(c)). 

Using a loss rate model incorporating vacuum loss, blackbody excitation, erasure conversion efficiency (estimated to be $\approx 90\%$), identification errors, and the internal state preparation fidelity, we obtain loss rates consistent with the measurements in Fig.~\ref{Fig_4}(c). We use this model and the measured lifetime to infer that $\approx80\%$ of blackbody errors are converted and detected correctly~\cite{Supplement}. With near-term technical improvements, efficiencies approaching $95\%$ should be possible~\cite{Supplement}. 

Having explored conversion of blackbody population errors followed by a single error detection image, we next explore mid-sequence detection. We measure the erasure-excised loss rate with five images equally spaced during a variable hold time and measure a loss rate of $\gamma_{p,E5}=0.12(1)\,\text{s}^{-1}$, which is lower than that obtained with one final erasure image by $0.014(15)\,\text{s}^{-1}$. This shows that the interspersed detection images do not lead to additional observable loss. In fact, the data hints that because mid-sequence detection reduces population leakage from $\mathcal{F}$ to higher vibrational and rotational states ($N\geq 5$), the lifetime of the detection states $\mathcal{F}$ is increased. We note that for these population loss investigations, we neither apply dynamical decoupling nor the spin-echo pulse in composite erasure detection, and are therefore free from the technical limitations of our Raman scheme (see Section~\ref{sec:techAndFundLimits}).

Lastly, we demonstrate that coherence loss due to blackbody leakage errors can be reduced by mid-circuit erasure conversion. Using a Ramsey sequence, we measure the hyperfine qubit coherence with and without erasure excision. For these measurements, we erasure-convert blackbody excitations mid-circuit, and perform error detection at the end of the sequence. During the Ramsey hold time, XY8 dynamical decoupling using the $X$-$B$ Raman configuration is applied. Without excision, the coherence loss rate is $1.64(4)\,\text{s}^{-1}$, consistent with the hyperfine qubit coherence in the $X$-$B$ Raman scheme, blackbody loss, and vacuum loss. In particular, off-resonant scattering from the Raman light contributes at least $\approx 50\%$ of the decoherence, and can be substantially reduced (see discussion in Section~\ref{sec:techAndFundLimits}). By excising data with erasures, we see a small but statistically significant ($\approx 3\sigma$) decrease in the decoherence rate of $0.17(6)\,\text{s}^{-1}$ (Fig.~\ref{Fig_4}(d)), consistent with the improvement the measured population loss $(0.162(15)\,\text{s}^{-1})$. Taken together, our measurements show that blackbody radiation primarily causes population loss but not qubit dephasing. Although the improvement is small compared to the total coherence loss rate, which is limited by population loss due to the XY8 Raman pulses, we expect that these losses can be substantially decreased by up to a factor of 60 to $0.1\,\text{s}^{-1}$ with a more optimal Raman light source (see Section~\ref{sec:techAndFundLimits}). At that point, blackbody radiation will become a non-negligible source of error that can be erasure-converted and detected.

\section{Conclusion and Outlook}
In summary, we have demonstrated 1) a site-resolved detection scheme that enables robust and record-level fidelities for preparing tweezers loaded with molecules occupying a single internal state, 2) an erasure detection scheme for molecules for the first time, with error rates at the few percent level, and 3) mid-circuit erasure conversion of blackbody induced leakage errors. These results open several possibilities. First, our achieved tweezer preparation fidelity of $\approx95\%$ opens access to quantum many-body simulation of spin models with few percent-level defects. Second, our erasure conversion scheme adds a powerful capability that could be useful for quantum error correction. Third, our work on erasure conversion of blackbody errors shows that they do not fundamentally impose an upper limit in circuit depth in molecular tweezer arrays, since they can be corrected mid-circuit when combined with tweezer reloading and state initialization~\cite{Singh2022RbCsAtomArray,Shaw2023DarkStateLoading}.

In addition to the possibilities above, our development of a new hyperfine qubit that is simultaneously coherent with a previously demonstrated rotational encoding opens the door to applications requiring 3-level systems, such as quantum information processing with qutrits, simulation of $S=1$ quantum spin models, bosonic $t$-$J$ models~\cite{homeier2023AFMtJ}, and lattice gauge theory models~\cite{Halimeh2023Rishon}. Separately, our work on detecting blackbody excitations could be relevant for investigations of molecule-based temperature standards~\cite{Norrgard2021QBBTherm}. Lastly, our work on enhanced tweezer preparation and catching blackbody errors could also aid molecular tweezer-based precision measurement experiments~\cite{Kozyryev2017PolyBSM,Anderegg2023trapCaOH}.

\begin{acknowledgments}
We thank Jeff Thompson, Waseem Bakr, and their groups for fruitful discussions. This work is supported by the National Science Foundation under Grant No. 2207518. C. M. H. acknowledges support from a Joseph Taylor Graduate Student Fellowship. S. J. L. and C. L. W. acknowledge support from Princeton Quantum Initiative Graduate Student Fellowships. L. W. C. acknowledges support from the Alfred P. Sloan Foundation under Grant No. FG-2022-19104.
\end{acknowledgments}

\bibliographystyle{apsrev4-1}

\begin{thebibliography}{54}%
\makeatletter
\providecommand \@ifxundefined [1]{%
 \@ifx{#1\undefined}
}%
\providecommand \@ifnum [1]{%
 \ifnum #1\expandafter \@firstoftwo
 \else \expandafter \@secondoftwo
 \fi
}%
\providecommand \@ifx [1]{%
 \ifx #1\expandafter \@firstoftwo
 \else \expandafter \@secondoftwo
 \fi
}%
\providecommand \natexlab [1]{#1}%
\providecommand \enquote  [1]{``#1''}%
\providecommand \bibnamefont  [1]{#1}%
\providecommand \bibfnamefont [1]{#1}%
\providecommand \citenamefont [1]{#1}%
\providecommand \href@noop [0]{\@secondoftwo}%
\providecommand \href [0]{\begingroup \@sanitize@url \@href}%
\providecommand \@href[1]{\@@startlink{#1}\@@href}%
\providecommand \@@href[1]{\endgroup#1\@@endlink}%
\providecommand \@sanitize@url [0]{\catcode `\\12\catcode `\$12\catcode
  `\&12\catcode `\#12\catcode `\^12\catcode `\_12\catcode `\%12\relax}%
\providecommand \@@startlink[1]{}%
\providecommand \@@endlink[0]{}%
\providecommand \url  [0]{\begingroup\@sanitize@url \@url }%
\providecommand \@url [1]{\endgroup\@href {#1}{\urlprefix }}%
\providecommand \urlprefix  [0]{URL }%
\providecommand \Eprint [0]{\href }%
\providecommand \doibase [0]{http://dx.doi.org/}%
\providecommand \selectlanguage [0]{\@gobble}%
\providecommand \bibinfo  [0]{\@secondoftwo}%
\providecommand \bibfield  [0]{\@secondoftwo}%
\providecommand \translation [1]{[#1]}%
\providecommand \BibitemOpen [0]{}%
\providecommand \bibitemStop [0]{}%
\providecommand \bibitemNoStop [0]{.\EOS\space}%
\providecommand \EOS [0]{\spacefactor3000\relax}%
\providecommand \BibitemShut  [1]{\csname bibitem#1\endcsname}%
\let\auto@bib@innerbib\@empty
\bibitem [{\citenamefont {Kaufman}\ and\ \citenamefont
  {Ni}(2021)}]{kaufman2021quantum}%
  \BibitemOpen
  \bibfield  {author} {\bibinfo {author} {\bibfnamefont {A.~M.}\ \bibnamefont
  {Kaufman}}\ and\ \bibinfo {author} {\bibfnamefont {K.-K.}\ \bibnamefont
  {Ni}},\ }\href {https://doi.org/10.1038/s41567-021-01357-2} {\bibfield
  {journal} {\bibinfo  {journal} {Nature Physics}\ }\textbf {\bibinfo {volume}
  {17}},\ \bibinfo {pages} {1324} (\bibinfo {year} {2021})}\BibitemShut
  {NoStop}%
\bibitem [{\citenamefont {Endres}\ \emph {et~al.}(2016)\citenamefont {Endres},
  \citenamefont {Bernien}, \citenamefont {Keesling}, \citenamefont {Levine},
  \citenamefont {Anschuetz}, \citenamefont {Krajenbrink}, \citenamefont
  {Senko}, \citenamefont {Vuleti\'{c}}, \citenamefont {Greiner},\ and\
  \citenamefont {Lukin}}]{Endres2016atomarray}%
  \BibitemOpen
  \bibfield  {author} {\bibinfo {author} {\bibfnamefont {M.}~\bibnamefont
  {Endres}}, \bibinfo {author} {\bibfnamefont {H.}~\bibnamefont {Bernien}},
  \bibinfo {author} {\bibfnamefont {A.}~\bibnamefont {Keesling}}, \bibinfo
  {author} {\bibfnamefont {H.}~\bibnamefont {Levine}}, \bibinfo {author}
  {\bibfnamefont {E.~R.}\ \bibnamefont {Anschuetz}}, \bibinfo {author}
  {\bibfnamefont {A.}~\bibnamefont {Krajenbrink}}, \bibinfo {author}
  {\bibfnamefont {C.}~\bibnamefont {Senko}}, \bibinfo {author} {\bibfnamefont
  {V.}~\bibnamefont {Vuleti\'{c}}}, \bibinfo {author} {\bibfnamefont
  {M.}~\bibnamefont {Greiner}}, \ and\ \bibinfo {author} {\bibfnamefont
  {M.~D.}\ \bibnamefont {Lukin}},\ }\href {\doibase 10.1126/science.aah3752}
  {\bibfield  {journal} {\bibinfo  {journal} {Science}\ }\textbf {\bibinfo
  {volume} {354}},\ \bibinfo {pages} {1024} (\bibinfo {year}
  {2016})}\BibitemShut {NoStop}%
\bibitem [{\citenamefont {Labuhn}\ \emph {et~al.}(2016)\citenamefont {Labuhn},
  \citenamefont {Barredo}, \citenamefont {Ravets}, \citenamefont
  {de~L\'{e}s\'{e}leuc}, \citenamefont {Macr\`{i}}, \citenamefont {Lahaye},\
  and\ \citenamefont {Browaeys}}]{Labuhn2016atomarray}%
  \BibitemOpen
  \bibfield  {author} {\bibinfo {author} {\bibfnamefont {H.}~\bibnamefont
  {Labuhn}}, \bibinfo {author} {\bibfnamefont {D.}~\bibnamefont {Barredo}},
  \bibinfo {author} {\bibfnamefont {S.}~\bibnamefont {Ravets}}, \bibinfo
  {author} {\bibfnamefont {S.}~\bibnamefont {de~L\'{e}s\'{e}leuc}}, \bibinfo
  {author} {\bibfnamefont {T.}~\bibnamefont {Macr\`{i}}}, \bibinfo {author}
  {\bibfnamefont {T.}~\bibnamefont {Lahaye}}, \ and\ \bibinfo {author}
  {\bibfnamefont {A.}~\bibnamefont {Browaeys}},\ }\href
  {https://doi.org/10.1038/nature18274} {\bibfield  {journal} {\bibinfo
  {journal} {Nature}\ }\textbf {\bibinfo {volume} {534}},\ \bibinfo {pages}
  {667} (\bibinfo {year} {2016})}\BibitemShut {NoStop}%
\bibitem [{\citenamefont {Yan}\ \emph {et~al.}(2013)\citenamefont {Yan},
  \citenamefont {Moses}, \citenamefont {Gadway}, \citenamefont {Covey},
  \citenamefont {Hazzard}, \citenamefont {Rey}, \citenamefont {Jin},\ and\
  \citenamefont {Ye}}]{yan2013observation}%
  \BibitemOpen
  \bibfield  {author} {\bibinfo {author} {\bibfnamefont {B.}~\bibnamefont
  {Yan}}, \bibinfo {author} {\bibfnamefont {S.~A.}\ \bibnamefont {Moses}},
  \bibinfo {author} {\bibfnamefont {B.}~\bibnamefont {Gadway}}, \bibinfo
  {author} {\bibfnamefont {J.~P.}\ \bibnamefont {Covey}}, \bibinfo {author}
  {\bibfnamefont {K.~R.}\ \bibnamefont {Hazzard}}, \bibinfo {author}
  {\bibfnamefont {A.~M.}\ \bibnamefont {Rey}}, \bibinfo {author} {\bibfnamefont
  {D.~S.}\ \bibnamefont {Jin}}, \ and\ \bibinfo {author} {\bibfnamefont
  {J.}~\bibnamefont {Ye}},\ }\href {https://doi.org/10.1038/nature12483}
  {\bibfield  {journal} {\bibinfo  {journal} {Nature}\ }\textbf {\bibinfo
  {volume} {501}},\ \bibinfo {pages} {521} (\bibinfo {year}
  {2013})}\BibitemShut {NoStop}%
\bibitem [{\citenamefont {Christakis}\ \emph {et~al.}(2023)\citenamefont
  {Christakis}, \citenamefont {Rosenberg}, \citenamefont {Raj}, \citenamefont
  {Chi}, \citenamefont {Morningstar}, \citenamefont {Huse}, \citenamefont
  {Yan},\ and\ \citenamefont {Bakr}}]{christakis2022probing}%
  \BibitemOpen
  \bibfield  {author} {\bibinfo {author} {\bibfnamefont {L.}~\bibnamefont
  {Christakis}}, \bibinfo {author} {\bibfnamefont {J.~S.}\ \bibnamefont
  {Rosenberg}}, \bibinfo {author} {\bibfnamefont {R.}~\bibnamefont {Raj}},
  \bibinfo {author} {\bibfnamefont {S.}~\bibnamefont {Chi}}, \bibinfo {author}
  {\bibfnamefont {A.}~\bibnamefont {Morningstar}}, \bibinfo {author}
  {\bibfnamefont {D.~A.}\ \bibnamefont {Huse}}, \bibinfo {author}
  {\bibfnamefont {Z.~Z.}\ \bibnamefont {Yan}}, \ and\ \bibinfo {author}
  {\bibfnamefont {W.~S.}\ \bibnamefont {Bakr}},\ }\href
  {https://doi.org/10.1038/s41586-022-05558-4} {\bibfield  {journal} {\bibinfo
  {journal} {Nature}\ }\textbf {\bibinfo {volume} {614}},\ \bibinfo {pages}
  {64} (\bibinfo {year} {2023})}\BibitemShut {NoStop}%
\bibitem [{\citenamefont {Gregory}\ \emph {et~al.}(2024)\citenamefont
  {Gregory}, \citenamefont {Fernley}, \citenamefont {Tao}, \citenamefont
  {Bromley}, \citenamefont {Stepp}, \citenamefont {Zhang}, \citenamefont
  {Kotochigova}, \citenamefont {Hazzard},\ and\ \citenamefont
  {Cornish}}]{Gregory2024second}%
  \BibitemOpen
  \bibfield  {author} {\bibinfo {author} {\bibfnamefont {P.~D.}\ \bibnamefont
  {Gregory}}, \bibinfo {author} {\bibfnamefont {L.~M.}\ \bibnamefont
  {Fernley}}, \bibinfo {author} {\bibfnamefont {A.~L.}\ \bibnamefont {Tao}},
  \bibinfo {author} {\bibfnamefont {S.~L.}\ \bibnamefont {Bromley}}, \bibinfo
  {author} {\bibfnamefont {J.}~\bibnamefont {Stepp}}, \bibinfo {author}
  {\bibfnamefont {Z.}~\bibnamefont {Zhang}}, \bibinfo {author} {\bibfnamefont
  {S.}~\bibnamefont {Kotochigova}}, \bibinfo {author} {\bibfnamefont
  {K.~R.~A.}\ \bibnamefont {Hazzard}}, \ and\ \bibinfo {author} {\bibfnamefont
  {S.~L.}\ \bibnamefont {Cornish}},\ }\href
  {https://doi.org/10.1038/s41567-023-02328-5} {\bibfield  {journal} {\bibinfo
  {journal} {Nature Physics}\ }\textbf {\bibinfo {volume} {20}},\ \bibinfo
  {pages} {415} (\bibinfo {year} {2024})}\BibitemShut {NoStop}%
\bibitem [{\citenamefont {Carr}\ \emph {et~al.}(2009)\citenamefont {Carr},
  \citenamefont {DeMille}, \citenamefont {Krems},\ and\ \citenamefont
  {Ye}}]{Carr2009review}%
  \BibitemOpen
  \bibfield  {author} {\bibinfo {author} {\bibfnamefont {L.~D.}\ \bibnamefont
  {Carr}}, \bibinfo {author} {\bibfnamefont {D.}~\bibnamefont {DeMille}},
  \bibinfo {author} {\bibfnamefont {R.~V.}\ \bibnamefont {Krems}}, \ and\
  \bibinfo {author} {\bibfnamefont {J.}~\bibnamefont {Ye}},\ }\href {\doibase
  10.1088/1367-2630/11/5/055049} {\bibfield  {journal} {\bibinfo  {journal}
  {New Journal of Physics}\ }\textbf {\bibinfo {volume} {11}},\ \bibinfo
  {pages} {055049} (\bibinfo {year} {2009})}\BibitemShut {NoStop}%
\bibitem [{\citenamefont {Bohn}\ \emph {et~al.}(2017)\citenamefont {Bohn},
  \citenamefont {Rey},\ and\ \citenamefont {Ye}}]{Bohn2017molreview}%
  \BibitemOpen
  \bibfield  {author} {\bibinfo {author} {\bibfnamefont {J.~L.}\ \bibnamefont
  {Bohn}}, \bibinfo {author} {\bibfnamefont {A.~M.}\ \bibnamefont {Rey}}, \
  and\ \bibinfo {author} {\bibfnamefont {J.}~\bibnamefont {Ye}},\ }\href
  {\doibase 10.1126/science.aam6299} {\bibfield  {journal} {\bibinfo  {journal}
  {Science}\ }\textbf {\bibinfo {volume} {357}},\ \bibinfo {pages} {1002}
  (\bibinfo {year} {2017})}\BibitemShut {NoStop}%
\bibitem [{\citenamefont {Safronova}\ \emph {et~al.}(2018)\citenamefont
  {Safronova}, \citenamefont {Budker}, \citenamefont {DeMille}, \citenamefont
  {Kimball}, \citenamefont {Derevianko},\ and\ \citenamefont
  {Clark}}]{Safronova2018NewPhysics}%
  \BibitemOpen
  \bibfield  {author} {\bibinfo {author} {\bibfnamefont {M.}~\bibnamefont
  {Safronova}}, \bibinfo {author} {\bibfnamefont {D.}~\bibnamefont {Budker}},
  \bibinfo {author} {\bibfnamefont {D.}~\bibnamefont {DeMille}}, \bibinfo
  {author} {\bibfnamefont {D.~F.~J.}\ \bibnamefont {Kimball}}, \bibinfo
  {author} {\bibfnamefont {A.}~\bibnamefont {Derevianko}}, \ and\ \bibinfo
  {author} {\bibfnamefont {C.~W.}\ \bibnamefont {Clark}},\ }\href
  {https://journals.aps.org/rmp/abstract/10.1103/RevModPhys.90.025008}
  {\bibfield  {journal} {\bibinfo  {journal} {Reviews of Modern Physics}\
  }\textbf {\bibinfo {volume} {90}},\ \bibinfo {pages} {025008} (\bibinfo
  {year} {2018})}\BibitemShut {NoStop}%
\bibitem [{\citenamefont {Kozyryev}\ and\ \citenamefont
  {Hutzler}(2017)}]{Kozyryev2017PolyBSM}%
  \BibitemOpen
  \bibfield  {author} {\bibinfo {author} {\bibfnamefont {I.}~\bibnamefont
  {Kozyryev}}\ and\ \bibinfo {author} {\bibfnamefont {N.~R.}\ \bibnamefont
  {Hutzler}},\ }\href
  {https://journals.aps.org/prl/abstract/10.1103/PhysRevLett.119.133002}
  {\bibfield  {journal} {\bibinfo  {journal} {Physical review letters}\
  }\textbf {\bibinfo {volume} {119}},\ \bibinfo {pages} {133002} (\bibinfo
  {year} {2017})}\BibitemShut {NoStop}%
\bibitem [{\citenamefont {Anderegg}\ \emph {et~al.}(2023)\citenamefont
  {Anderegg}, \citenamefont {Vilas}, \citenamefont {Hallas}, \citenamefont
  {Robichaud}, \citenamefont {Jadbabaie}, \citenamefont {Doyle},\ and\
  \citenamefont {Hutzler}}]{Anderegg2023trapCaOH}%
  \BibitemOpen
  \bibfield  {author} {\bibinfo {author} {\bibfnamefont {L.}~\bibnamefont
  {Anderegg}}, \bibinfo {author} {\bibfnamefont {N.~B.}\ \bibnamefont {Vilas}},
  \bibinfo {author} {\bibfnamefont {C.}~\bibnamefont {Hallas}}, \bibinfo
  {author} {\bibfnamefont {P.}~\bibnamefont {Robichaud}}, \bibinfo {author}
  {\bibfnamefont {A.}~\bibnamefont {Jadbabaie}}, \bibinfo {author}
  {\bibfnamefont {J.~M.}\ \bibnamefont {Doyle}}, \ and\ \bibinfo {author}
  {\bibfnamefont {N.~R.}\ \bibnamefont {Hutzler}},\ }\href {\doibase
  10.1126/science.adg8155} {\bibfield  {journal} {\bibinfo  {journal}
  {Science}\ }\textbf {\bibinfo {volume} {382}},\ \bibinfo {pages} {665}
  (\bibinfo {year} {2023})}\BibitemShut {NoStop}%
\bibitem [{\citenamefont {DeMille}(2002)}]{demille2002quantum}%
  \BibitemOpen
  \bibfield  {author} {\bibinfo {author} {\bibfnamefont {D.}~\bibnamefont
  {DeMille}},\ }\href {https://doi.org/10.1103/PhysRevLett.88.067901}
  {\bibfield  {journal} {\bibinfo  {journal} {Physical Review Letters}\
  }\textbf {\bibinfo {volume} {88}},\ \bibinfo {pages} {067901} (\bibinfo
  {year} {2002})}\BibitemShut {NoStop}%
\bibitem [{\citenamefont {Ni}\ \emph {et~al.}(2018)\citenamefont {Ni},
  \citenamefont {Rosenband},\ and\ \citenamefont {Grimes}}]{Ni2018gate}%
  \BibitemOpen
  \bibfield  {author} {\bibinfo {author} {\bibfnamefont {K.-K.}\ \bibnamefont
  {Ni}}, \bibinfo {author} {\bibfnamefont {T.}~\bibnamefont {Rosenband}}, \
  and\ \bibinfo {author} {\bibfnamefont {D.~D.}\ \bibnamefont {Grimes}},\
  }\href {\doibase 10.1039/C8SC02355G} {\bibfield  {journal} {\bibinfo
  {journal} {Chem. Sci.}\ }\textbf {\bibinfo {volume} {9}},\ \bibinfo {pages}
  {6830} (\bibinfo {year} {2018})}\BibitemShut {NoStop}%
\bibitem [{\citenamefont {Yu}\ \emph {et~al.}(2019)\citenamefont {Yu},
  \citenamefont {Cheuk}, \citenamefont {Kozyryev},\ and\ \citenamefont
  {Doyle}}]{Yu2019symtop}%
  \BibitemOpen
  \bibfield  {author} {\bibinfo {author} {\bibfnamefont {P.}~\bibnamefont
  {Yu}}, \bibinfo {author} {\bibfnamefont {L.~W.}\ \bibnamefont {Cheuk}},
  \bibinfo {author} {\bibfnamefont {I.}~\bibnamefont {Kozyryev}}, \ and\
  \bibinfo {author} {\bibfnamefont {J.~M.}\ \bibnamefont {Doyle}},\ }\href
  {https://iopscience.iop.org/article/10.1088/1367-2630/ab428d} {\bibfield
  {journal} {\bibinfo  {journal} {New Journal of Physics}\ }\textbf {\bibinfo
  {volume} {21}},\ \bibinfo {pages} {093049} (\bibinfo {year}
  {2019})}\BibitemShut {NoStop}%
\bibitem [{\citenamefont {Micheli}\ \emph {et~al.}(2006)\citenamefont
  {Micheli}, \citenamefont {Brennen},\ and\ \citenamefont
  {Zoller}}]{Micheli2006spintoolbox}%
  \BibitemOpen
  \bibfield  {author} {\bibinfo {author} {\bibfnamefont {A.}~\bibnamefont
  {Micheli}}, \bibinfo {author} {\bibfnamefont {G.~K.}\ \bibnamefont
  {Brennen}}, \ and\ \bibinfo {author} {\bibfnamefont {P.}~\bibnamefont
  {Zoller}},\ }\href {\doibase 10.1038/nphys287} {\bibfield  {journal}
  {\bibinfo  {journal} {Nature Physics}\ }\textbf {\bibinfo {volume} {2}},\
  \bibinfo {pages} {341} (\bibinfo {year} {2006})}\BibitemShut {NoStop}%
\bibitem [{\citenamefont {Gadway}\ and\ \citenamefont
  {Yan}(2016)}]{Gadway2016StronglyInteracting}%
  \BibitemOpen
  \bibfield  {author} {\bibinfo {author} {\bibfnamefont {B.}~\bibnamefont
  {Gadway}}\ and\ \bibinfo {author} {\bibfnamefont {B.}~\bibnamefont {Yan}},\
  }\href {https://iopscience.iop.org/article/10.1088/0953-4075/49/15/152002}
  {\bibfield  {journal} {\bibinfo  {journal} {Journal of Physics B: Atomic,
  Molecular and Optical Physics}\ }\textbf {\bibinfo {volume} {49}},\ \bibinfo
  {pages} {152002} (\bibinfo {year} {2016})}\BibitemShut {NoStop}%
\bibitem [{\citenamefont {Anderegg}\ \emph {et~al.}(2019)\citenamefont
  {Anderegg}, \citenamefont {Cheuk}, \citenamefont {Bao}, \citenamefont
  {Burchesky}, \citenamefont {Ketterle}, \citenamefont {Ni},\ and\
  \citenamefont {Doyle}}]{Anderegg2019Tweezer}%
  \BibitemOpen
  \bibfield  {author} {\bibinfo {author} {\bibfnamefont {L.}~\bibnamefont
  {Anderegg}}, \bibinfo {author} {\bibfnamefont {L.~W.}\ \bibnamefont {Cheuk}},
  \bibinfo {author} {\bibfnamefont {Y.}~\bibnamefont {Bao}}, \bibinfo {author}
  {\bibfnamefont {S.}~\bibnamefont {Burchesky}}, \bibinfo {author}
  {\bibfnamefont {W.}~\bibnamefont {Ketterle}}, \bibinfo {author}
  {\bibfnamefont {K.-K.}\ \bibnamefont {Ni}}, \ and\ \bibinfo {author}
  {\bibfnamefont {J.~M.}\ \bibnamefont {Doyle}},\ }\href {\doibase
  10.1126/science.aax1265} {\bibfield  {journal} {\bibinfo  {journal}
  {Science}\ }\textbf {\bibinfo {volume} {365}},\ \bibinfo {pages} {1156}
  (\bibinfo {year} {2019})}\BibitemShut {NoStop}%
\bibitem [{\citenamefont {Zhang}\ \emph {et~al.}(2022)\citenamefont {Zhang},
  \citenamefont {Picard}, \citenamefont {Cairncross}, \citenamefont {Wang},
  \citenamefont {Yu}, \citenamefont {Fang},\ and\ \citenamefont
  {Ni}}]{Zhang2021NaCsArray}%
  \BibitemOpen
  \bibfield  {author} {\bibinfo {author} {\bibfnamefont {J.~T.}\ \bibnamefont
  {Zhang}}, \bibinfo {author} {\bibfnamefont {L.~R.~B.}\ \bibnamefont
  {Picard}}, \bibinfo {author} {\bibfnamefont {W.~B.}\ \bibnamefont
  {Cairncross}}, \bibinfo {author} {\bibfnamefont {K.}~\bibnamefont {Wang}},
  \bibinfo {author} {\bibfnamefont {Y.}~\bibnamefont {Yu}}, \bibinfo {author}
  {\bibfnamefont {F.}~\bibnamefont {Fang}}, \ and\ \bibinfo {author}
  {\bibfnamefont {K.-K.}\ \bibnamefont {Ni}},\ }\href {\doibase
  10.1088/2058-9565/ac676c} {\bibfield  {journal} {\bibinfo  {journal} {Quantum
  Science and Technology}\ }\textbf {\bibinfo {volume} {7}},\ \bibinfo {pages}
  {035006} (\bibinfo {year} {2022})}\BibitemShut {NoStop}%
\bibitem [{\citenamefont {Holland}\ \emph
  {et~al.}(2023{\natexlab{a}})\citenamefont {Holland}, \citenamefont {Lu},\
  and\ \citenamefont {Cheuk}}]{Holland2023bichromatic}%
  \BibitemOpen
  \bibfield  {author} {\bibinfo {author} {\bibfnamefont {C.~M.}\ \bibnamefont
  {Holland}}, \bibinfo {author} {\bibfnamefont {Y.}~\bibnamefont {Lu}}, \ and\
  \bibinfo {author} {\bibfnamefont {L.~W.}\ \bibnamefont {Cheuk}},\ }\href
  {\doibase 10.1103/PhysRevLett.131.053202} {\bibfield  {journal} {\bibinfo
  {journal} {Phys. Rev. Lett.}\ }\textbf {\bibinfo {volume} {131}},\ \bibinfo
  {pages} {053202} (\bibinfo {year} {2023}{\natexlab{a}})}\BibitemShut
  {NoStop}%
\bibitem [{\citenamefont {Ruttley}\ \emph {et~al.}(2023)\citenamefont
  {Ruttley}, \citenamefont {Guttridge}, \citenamefont {Spence}, \citenamefont
  {Bird}, \citenamefont {Le~Sueur}, \citenamefont {Hutson},\ and\ \citenamefont
  {Cornish}}]{Ruttley2023RbCsTweezer}%
  \BibitemOpen
  \bibfield  {author} {\bibinfo {author} {\bibfnamefont {D.~K.}\ \bibnamefont
  {Ruttley}}, \bibinfo {author} {\bibfnamefont {A.}~\bibnamefont {Guttridge}},
  \bibinfo {author} {\bibfnamefont {S.}~\bibnamefont {Spence}}, \bibinfo
  {author} {\bibfnamefont {R.~C.}\ \bibnamefont {Bird}}, \bibinfo {author}
  {\bibfnamefont {C.~R.}\ \bibnamefont {Le~Sueur}}, \bibinfo {author}
  {\bibfnamefont {J.~M.}\ \bibnamefont {Hutson}}, \ and\ \bibinfo {author}
  {\bibfnamefont {S.~L.}\ \bibnamefont {Cornish}},\ }\href {\doibase
  10.1103/PhysRevLett.130.223401} {\bibfield  {journal} {\bibinfo  {journal}
  {Phys. Rev. Lett.}\ }\textbf {\bibinfo {volume} {130}},\ \bibinfo {pages}
  {223401} (\bibinfo {year} {2023})}\BibitemShut {NoStop}%
\bibitem [{\citenamefont {Vilas}\ \emph {et~al.}(2024)\citenamefont {Vilas},
  \citenamefont {Robichaud}, \citenamefont {Hallas}, \citenamefont {Li},
  \citenamefont {Anderegg},\ and\ \citenamefont
  {Doyle}}]{Vilas2023CaOHTweezer}%
  \BibitemOpen
  \bibfield  {author} {\bibinfo {author} {\bibfnamefont {N.~B.}\ \bibnamefont
  {Vilas}}, \bibinfo {author} {\bibfnamefont {P.}~\bibnamefont {Robichaud}},
  \bibinfo {author} {\bibfnamefont {C.}~\bibnamefont {Hallas}}, \bibinfo
  {author} {\bibfnamefont {G.~K.}\ \bibnamefont {Li}}, \bibinfo {author}
  {\bibfnamefont {L.}~\bibnamefont {Anderegg}}, \ and\ \bibinfo {author}
  {\bibfnamefont {J.~M.}\ \bibnamefont {Doyle}},\ }\href
  {https://doi.org/10.1038/s41586-024-07199-1} {\bibfield  {journal} {\bibinfo
  {journal} {Nature}\ }\textbf {\bibinfo {volume} {628}},\ \bibinfo {pages}
  {282} (\bibinfo {year} {2024})}\BibitemShut {NoStop}%
\bibitem [{\citenamefont {Burchesky}\ \emph {et~al.}(2021)\citenamefont
  {Burchesky}, \citenamefont {Anderegg}, \citenamefont {Bao}, \citenamefont
  {Yu}, \citenamefont {Chae}, \citenamefont {Ketterle}, \citenamefont {Ni},\
  and\ \citenamefont {Doyle}}]{burchesky2021rotcoh}%
  \BibitemOpen
  \bibfield  {author} {\bibinfo {author} {\bibfnamefont {S.}~\bibnamefont
  {Burchesky}}, \bibinfo {author} {\bibfnamefont {L.}~\bibnamefont {Anderegg}},
  \bibinfo {author} {\bibfnamefont {Y.}~\bibnamefont {Bao}}, \bibinfo {author}
  {\bibfnamefont {S.~S.}\ \bibnamefont {Yu}}, \bibinfo {author} {\bibfnamefont
  {E.}~\bibnamefont {Chae}}, \bibinfo {author} {\bibfnamefont {W.}~\bibnamefont
  {Ketterle}}, \bibinfo {author} {\bibfnamefont {K.-K.}\ \bibnamefont {Ni}}, \
  and\ \bibinfo {author} {\bibfnamefont {J.~M.}\ \bibnamefont {Doyle}},\ }\href
  {\doibase 10.1103/PhysRevLett.127.123202} {\bibfield  {journal} {\bibinfo
  {journal} {Phys. Rev. Lett.}\ }\textbf {\bibinfo {volume} {127}},\ \bibinfo
  {pages} {123202} (\bibinfo {year} {2021})}\BibitemShut {NoStop}%
\bibitem [{\citenamefont {Park}\ \emph {et~al.}(2023)\citenamefont {Park},
  \citenamefont {Picard}, \citenamefont {Patenotte}, \citenamefont {Zhang},
  \citenamefont {Rosenband},\ and\ \citenamefont {Ni}}]{Park2023rotcoh}%
  \BibitemOpen
  \bibfield  {author} {\bibinfo {author} {\bibfnamefont {A.~J.}\ \bibnamefont
  {Park}}, \bibinfo {author} {\bibfnamefont {L.~R.~B.}\ \bibnamefont {Picard}},
  \bibinfo {author} {\bibfnamefont {G.~E.}\ \bibnamefont {Patenotte}}, \bibinfo
  {author} {\bibfnamefont {J.~T.}\ \bibnamefont {Zhang}}, \bibinfo {author}
  {\bibfnamefont {T.}~\bibnamefont {Rosenband}}, \ and\ \bibinfo {author}
  {\bibfnamefont {K.-K.}\ \bibnamefont {Ni}},\ }\href {\doibase
  10.1103/PhysRevLett.131.183401} {\bibfield  {journal} {\bibinfo  {journal}
  {Phys. Rev. Lett.}\ }\textbf {\bibinfo {volume} {131}},\ \bibinfo {pages}
  {183401} (\bibinfo {year} {2023})}\BibitemShut {NoStop}%
\bibitem [{\citenamefont {Holland}\ \emph
  {et~al.}(2023{\natexlab{b}})\citenamefont {Holland}, \citenamefont {Lu},\
  and\ \citenamefont {Cheuk}}]{Holland2023Entanglement}%
  \BibitemOpen
  \bibfield  {author} {\bibinfo {author} {\bibfnamefont {C.~M.}\ \bibnamefont
  {Holland}}, \bibinfo {author} {\bibfnamefont {Y.}~\bibnamefont {Lu}}, \ and\
  \bibinfo {author} {\bibfnamefont {L.~W.}\ \bibnamefont {Cheuk}},\ }\href
  {\doibase 10.1126/science.adf4272} {\bibfield  {journal} {\bibinfo  {journal}
  {Science}\ }\textbf {\bibinfo {volume} {382}},\ \bibinfo {pages} {1143}
  (\bibinfo {year} {2023}{\natexlab{b}})}\BibitemShut {NoStop}%
\bibitem [{\citenamefont {Bao}\ \emph {et~al.}(2023{\natexlab{a}})\citenamefont
  {Bao}, \citenamefont {Yu}, \citenamefont {Anderegg}, \citenamefont {Chae},
  \citenamefont {Ketterle}, \citenamefont {Ni},\ and\ \citenamefont
  {Doyle}}]{Bao2023Entanglement}%
  \BibitemOpen
  \bibfield  {author} {\bibinfo {author} {\bibfnamefont {Y.}~\bibnamefont
  {Bao}}, \bibinfo {author} {\bibfnamefont {S.~S.}\ \bibnamefont {Yu}},
  \bibinfo {author} {\bibfnamefont {L.}~\bibnamefont {Anderegg}}, \bibinfo
  {author} {\bibfnamefont {E.}~\bibnamefont {Chae}}, \bibinfo {author}
  {\bibfnamefont {W.}~\bibnamefont {Ketterle}}, \bibinfo {author}
  {\bibfnamefont {K.-K.}\ \bibnamefont {Ni}}, \ and\ \bibinfo {author}
  {\bibfnamefont {J.~M.}\ \bibnamefont {Doyle}},\ }\href {\doibase
  10.1126/science.adf8999} {\bibfield  {journal} {\bibinfo  {journal}
  {Science}\ }\textbf {\bibinfo {volume} {382}},\ \bibinfo {pages} {1138}
  (\bibinfo {year} {2023}{\natexlab{a}})}\BibitemShut {NoStop}%
\bibitem [{\citenamefont {Lu}\ \emph {et~al.}(2024)\citenamefont {Lu},
  \citenamefont {Li}, \citenamefont {Holland},\ and\ \citenamefont
  {Cheuk}}]{Lu2024RSC}%
  \BibitemOpen
  \bibfield  {author} {\bibinfo {author} {\bibfnamefont {Y.}~\bibnamefont
  {Lu}}, \bibinfo {author} {\bibfnamefont {S.~J.}\ \bibnamefont {Li}}, \bibinfo
  {author} {\bibfnamefont {C.~M.}\ \bibnamefont {Holland}}, \ and\ \bibinfo
  {author} {\bibfnamefont {L.~W.}\ \bibnamefont {Cheuk}},\ }\href
  {https://www.nature.com/articles/s41567-023-02346-3} {\bibfield  {journal}
  {\bibinfo  {journal} {Nature Physics}\ ,\ \bibinfo {pages} {1}} (\bibinfo
  {year} {2024})}\BibitemShut {NoStop}%
\bibitem [{\citenamefont {Bao}\ \emph {et~al.}(2023{\natexlab{b}})\citenamefont
  {Bao}, \citenamefont {Yu}, \citenamefont {You}, \citenamefont {Anderegg},
  \citenamefont {Chae}, \citenamefont {Ketterle}, \citenamefont {Ni},\ and\
  \citenamefont {Doyle}}]{Bao2023RSC}%
  \BibitemOpen
  \bibfield  {author} {\bibinfo {author} {\bibfnamefont {Y.}~\bibnamefont
  {Bao}}, \bibinfo {author} {\bibfnamefont {S.~S.}\ \bibnamefont {Yu}},
  \bibinfo {author} {\bibfnamefont {J.}~\bibnamefont {You}}, \bibinfo {author}
  {\bibfnamefont {L.}~\bibnamefont {Anderegg}}, \bibinfo {author}
  {\bibfnamefont {E.}~\bibnamefont {Chae}}, \bibinfo {author} {\bibfnamefont
  {W.}~\bibnamefont {Ketterle}}, \bibinfo {author} {\bibfnamefont {K.-K.}\
  \bibnamefont {Ni}}, \ and\ \bibinfo {author} {\bibfnamefont {J.~M.}\
  \bibnamefont {Doyle}},\ }\href@noop {} {\bibfield  {journal} {\bibinfo
  {journal} {arXiv:2309.08706v1}\ } (\bibinfo {year}
  {2023}{\natexlab{b}})}\BibitemShut {NoStop}%
\bibitem [{\citenamefont {Scholl}\ \emph
  {et~al.}(2023{\natexlab{a}})\citenamefont {Scholl}, \citenamefont {Shaw},
  \citenamefont {Tsai}, \citenamefont {Finkelstein}, \citenamefont {Choi},\
  and\ \citenamefont {Endres}}]{Scholl2023Erasure}%
  \BibitemOpen
  \bibfield  {author} {\bibinfo {author} {\bibfnamefont {P.}~\bibnamefont
  {Scholl}}, \bibinfo {author} {\bibfnamefont {A.~L.}\ \bibnamefont {Shaw}},
  \bibinfo {author} {\bibfnamefont {R.~B.-S.}\ \bibnamefont {Tsai}}, \bibinfo
  {author} {\bibfnamefont {R.}~\bibnamefont {Finkelstein}}, \bibinfo {author}
  {\bibfnamefont {J.}~\bibnamefont {Choi}}, \ and\ \bibinfo {author}
  {\bibfnamefont {M.}~\bibnamefont {Endres}},\ }\href
  {https://www.nature.com/articles/s41586-023-06516-4} {\bibfield  {journal}
  {\bibinfo  {journal} {Nature}\ }\textbf {\bibinfo {volume} {622}},\ \bibinfo
  {pages} {273} (\bibinfo {year} {2023}{\natexlab{a}})}\BibitemShut {NoStop}%
\bibitem [{\citenamefont {Ma}\ \emph {et~al.}(2023)\citenamefont {Ma},
  \citenamefont {Liu}, \citenamefont {Peng}, \citenamefont {Zhang},
  \citenamefont {Jandura}, \citenamefont {Claes}, \citenamefont {Burgers},
  \citenamefont {Pupillo}, \citenamefont {Puri},\ and\ \citenamefont
  {Thompson}}]{Ma2023Erasure}%
  \BibitemOpen
  \bibfield  {author} {\bibinfo {author} {\bibfnamefont {S.}~\bibnamefont
  {Ma}}, \bibinfo {author} {\bibfnamefont {G.}~\bibnamefont {Liu}}, \bibinfo
  {author} {\bibfnamefont {P.}~\bibnamefont {Peng}}, \bibinfo {author}
  {\bibfnamefont {B.}~\bibnamefont {Zhang}}, \bibinfo {author} {\bibfnamefont
  {S.}~\bibnamefont {Jandura}}, \bibinfo {author} {\bibfnamefont
  {J.}~\bibnamefont {Claes}}, \bibinfo {author} {\bibfnamefont {A.~P.}\
  \bibnamefont {Burgers}}, \bibinfo {author} {\bibfnamefont {G.}~\bibnamefont
  {Pupillo}}, \bibinfo {author} {\bibfnamefont {S.}~\bibnamefont {Puri}}, \
  and\ \bibinfo {author} {\bibfnamefont {J.~D.}\ \bibnamefont {Thompson}},\
  }\href {https://www.nature.com/articles/s41586-023-06438-1} {\bibfield
  {journal} {\bibinfo  {journal} {Nature}\ }\textbf {\bibinfo {volume} {622}},\
  \bibinfo {pages} {279} (\bibinfo {year} {2023})}\BibitemShut {NoStop}%
\bibitem [{\citenamefont {Scholl}\ \emph
  {et~al.}(2023{\natexlab{b}})\citenamefont {Scholl}, \citenamefont {Shaw},
  \citenamefont {Finkelstein}, \citenamefont {Tsai}, \citenamefont {Choi},\
  and\ \citenamefont {Endres}}]{Scholl2023HyperEntanglement}%
  \BibitemOpen
  \bibfield  {author} {\bibinfo {author} {\bibfnamefont {P.}~\bibnamefont
  {Scholl}}, \bibinfo {author} {\bibfnamefont {A.~L.}\ \bibnamefont {Shaw}},
  \bibinfo {author} {\bibfnamefont {R.}~\bibnamefont {Finkelstein}}, \bibinfo
  {author} {\bibfnamefont {R.~B.-S.}\ \bibnamefont {Tsai}}, \bibinfo {author}
  {\bibfnamefont {J.}~\bibnamefont {Choi}}, \ and\ \bibinfo {author}
  {\bibfnamefont {M.}~\bibnamefont {Endres}},\ }\href@noop {} {\bibfield
  {journal} {\bibinfo  {journal} {arXiv:2311.15580}\ } (\bibinfo {year}
  {2023}{\natexlab{b}})}\BibitemShut {NoStop}%
\bibitem [{\citenamefont {Picard}\ \emph {et~al.}(2024)\citenamefont {Picard},
  \citenamefont {Patenotte}, \citenamefont {Park}, \citenamefont
  {Gebretsadkan},\ and\ \citenamefont {Ni}}]{Picard2024siteselective}%
  \BibitemOpen
  \bibfield  {author} {\bibinfo {author} {\bibfnamefont {L.~R.~B.}\
  \bibnamefont {Picard}}, \bibinfo {author} {\bibfnamefont {G.~E.}\
  \bibnamefont {Patenotte}}, \bibinfo {author} {\bibfnamefont {A.~J.}\
  \bibnamefont {Park}}, \bibinfo {author} {\bibfnamefont {S.~F.}\ \bibnamefont
  {Gebretsadkan}}, \ and\ \bibinfo {author} {\bibfnamefont {K.-K.}\
  \bibnamefont {Ni}},\ }\href {\doibase 10.1103/PRXQuantum.5.020344} {\bibfield
   {journal} {\bibinfo  {journal} {PRX Quantum}\ }\textbf {\bibinfo {volume}
  {5}},\ \bibinfo {pages} {020344} (\bibinfo {year} {2024})}\BibitemShut
  {NoStop}%
\bibitem [{\citenamefont {Ruttley}\ \emph {et~al.}(2024)\citenamefont
  {Ruttley}, \citenamefont {Guttridge}, \citenamefont {Hepworth},\ and\
  \citenamefont {Cornish}}]{Ruttley2024feedback}%
  \BibitemOpen
  \bibfield  {author} {\bibinfo {author} {\bibfnamefont {D.~K.}\ \bibnamefont
  {Ruttley}}, \bibinfo {author} {\bibfnamefont {A.}~\bibnamefont {Guttridge}},
  \bibinfo {author} {\bibfnamefont {T.~R.}\ \bibnamefont {Hepworth}}, \ and\
  \bibinfo {author} {\bibfnamefont {S.~L.}\ \bibnamefont {Cornish}},\ }\href
  {\doibase 10.1103/PRXQuantum.5.020333} {\bibfield  {journal} {\bibinfo
  {journal} {PRX Quantum}\ }\textbf {\bibinfo {volume} {5}},\ \bibinfo {pages}
  {020333} (\bibinfo {year} {2024})}\BibitemShut {NoStop}%
\bibitem [{\citenamefont {Grassl}\ \emph {et~al.}(1997)\citenamefont {Grassl},
  \citenamefont {Beth},\ and\ \citenamefont
  {Pellizzari}}]{Grassl1997erasurecodes}%
  \BibitemOpen
  \bibfield  {author} {\bibinfo {author} {\bibfnamefont {M.}~\bibnamefont
  {Grassl}}, \bibinfo {author} {\bibfnamefont {T.}~\bibnamefont {Beth}}, \ and\
  \bibinfo {author} {\bibfnamefont {T.}~\bibnamefont {Pellizzari}},\ }\href
  {\doibase 10.1103/PhysRevA.56.33} {\bibfield  {journal} {\bibinfo  {journal}
  {Phys. Rev. A}\ }\textbf {\bibinfo {volume} {56}},\ \bibinfo {pages} {33}
  (\bibinfo {year} {1997})}\BibitemShut {NoStop}%
\bibitem [{\citenamefont {Bennett}\ \emph {et~al.}(1997)\citenamefont
  {Bennett}, \citenamefont {DiVincenzo},\ and\ \citenamefont
  {Smolin}}]{Bennett1997ErasureCapacity}%
  \BibitemOpen
  \bibfield  {author} {\bibinfo {author} {\bibfnamefont {C.~H.}\ \bibnamefont
  {Bennett}}, \bibinfo {author} {\bibfnamefont {D.~P.}\ \bibnamefont
  {DiVincenzo}}, \ and\ \bibinfo {author} {\bibfnamefont {J.~A.}\ \bibnamefont
  {Smolin}},\ }\href {\doibase 10.1103/PhysRevLett.78.3217} {\bibfield
  {journal} {\bibinfo  {journal} {Phys. Rev. Lett.}\ }\textbf {\bibinfo
  {volume} {78}},\ \bibinfo {pages} {3217} (\bibinfo {year}
  {1997})}\BibitemShut {NoStop}%
\bibitem [{\citenamefont {Wu}\ \emph {et~al.}(2022)\citenamefont {Wu},
  \citenamefont {Kolkowitz}, \citenamefont {Puri},\ and\ \citenamefont
  {Thompson}}]{Wu2022Erasure}%
  \BibitemOpen
  \bibfield  {author} {\bibinfo {author} {\bibfnamefont {Y.}~\bibnamefont
  {Wu}}, \bibinfo {author} {\bibfnamefont {S.}~\bibnamefont {Kolkowitz}},
  \bibinfo {author} {\bibfnamefont {S.}~\bibnamefont {Puri}}, \ and\ \bibinfo
  {author} {\bibfnamefont {J.~D.}\ \bibnamefont {Thompson}},\ }\href
  {https://www.nature.com/articles/s41467-022-32094-6} {\bibfield  {journal}
  {\bibinfo  {journal} {Nature communications}\ }\textbf {\bibinfo {volume}
  {13}},\ \bibinfo {pages} {4657} (\bibinfo {year} {2022})}\BibitemShut
  {NoStop}%
\bibitem [{\citenamefont {Kubica}\ \emph {et~al.}(2023)\citenamefont {Kubica},
  \citenamefont {Haim}, \citenamefont {Vaknin}, \citenamefont {Levine},
  \citenamefont {Brand\~ao},\ and\ \citenamefont
  {Retzker}}]{Kubica2023erasureproposal}%
  \BibitemOpen
  \bibfield  {author} {\bibinfo {author} {\bibfnamefont {A.}~\bibnamefont
  {Kubica}}, \bibinfo {author} {\bibfnamefont {A.}~\bibnamefont {Haim}},
  \bibinfo {author} {\bibfnamefont {Y.}~\bibnamefont {Vaknin}}, \bibinfo
  {author} {\bibfnamefont {H.}~\bibnamefont {Levine}}, \bibinfo {author}
  {\bibfnamefont {F.}~\bibnamefont {Brand\~ao}}, \ and\ \bibinfo {author}
  {\bibfnamefont {A.}~\bibnamefont {Retzker}},\ }\href {\doibase
  10.1103/PhysRevX.13.041022} {\bibfield  {journal} {\bibinfo  {journal} {Phys.
  Rev. X}\ }\textbf {\bibinfo {volume} {13}},\ \bibinfo {pages} {041022}
  (\bibinfo {year} {2023})}\BibitemShut {NoStop}%
\bibitem [{\citenamefont {Kang}\ \emph {et~al.}(2023)\citenamefont {Kang},
  \citenamefont {Campbell},\ and\ \citenamefont
  {Brown}}]{Kang2023erasureproposal}%
  \BibitemOpen
  \bibfield  {author} {\bibinfo {author} {\bibfnamefont {M.}~\bibnamefont
  {Kang}}, \bibinfo {author} {\bibfnamefont {W.~C.}\ \bibnamefont {Campbell}},
  \ and\ \bibinfo {author} {\bibfnamefont {K.~R.}\ \bibnamefont {Brown}},\
  }\href {\doibase 10.1103/PRXQuantum.4.020358} {\bibfield  {journal} {\bibinfo
   {journal} {PRX Quantum}\ }\textbf {\bibinfo {volume} {4}},\ \bibinfo {pages}
  {020358} (\bibinfo {year} {2023})}\BibitemShut {NoStop}%
\bibitem [{\citenamefont {Teoh}\ \emph {et~al.}(2023)\citenamefont {Teoh},
  \citenamefont {Winkel}, \citenamefont {Babla}, \citenamefont {Chapman},
  \citenamefont {Claes}, \citenamefont {de~Graaf}, \citenamefont {Garmon},
  \citenamefont {Kalfus}, \citenamefont {Lu}, \citenamefont {Maiti},
  \citenamefont {Sahay}, \citenamefont {Thakur}, \citenamefont {Tsunoda},
  \citenamefont {Xue}, \citenamefont {Frunzio}, \citenamefont {Girvin},
  \citenamefont {Puri},\ and\ \citenamefont
  {Schoelkopf}}]{Teoh2023erasureproposal}%
  \BibitemOpen
  \bibfield  {author} {\bibinfo {author} {\bibfnamefont {J.~D.}\ \bibnamefont
  {Teoh}}, \bibinfo {author} {\bibfnamefont {P.}~\bibnamefont {Winkel}},
  \bibinfo {author} {\bibfnamefont {H.~K.}\ \bibnamefont {Babla}}, \bibinfo
  {author} {\bibfnamefont {B.~J.}\ \bibnamefont {Chapman}}, \bibinfo {author}
  {\bibfnamefont {J.}~\bibnamefont {Claes}}, \bibinfo {author} {\bibfnamefont
  {S.~J.}\ \bibnamefont {de~Graaf}}, \bibinfo {author} {\bibfnamefont
  {J.~W.~O.}\ \bibnamefont {Garmon}}, \bibinfo {author} {\bibfnamefont {W.~D.}\
  \bibnamefont {Kalfus}}, \bibinfo {author} {\bibfnamefont {Y.}~\bibnamefont
  {Lu}}, \bibinfo {author} {\bibfnamefont {A.}~\bibnamefont {Maiti}}, \bibinfo
  {author} {\bibfnamefont {K.}~\bibnamefont {Sahay}}, \bibinfo {author}
  {\bibfnamefont {N.}~\bibnamefont {Thakur}}, \bibinfo {author} {\bibfnamefont
  {T.}~\bibnamefont {Tsunoda}}, \bibinfo {author} {\bibfnamefont {S.~H.}\
  \bibnamefont {Xue}}, \bibinfo {author} {\bibfnamefont {L.}~\bibnamefont
  {Frunzio}}, \bibinfo {author} {\bibfnamefont {S.~M.}\ \bibnamefont {Girvin}},
  \bibinfo {author} {\bibfnamefont {S.}~\bibnamefont {Puri}}, \ and\ \bibinfo
  {author} {\bibfnamefont {R.~J.}\ \bibnamefont {Schoelkopf}},\ }\href
  {\doibase 10.1073/pnas.2221736120} {\bibfield  {journal} {\bibinfo  {journal}
  {Proceedings of the National Academy of Sciences}\ }\textbf {\bibinfo
  {volume} {120}},\ \bibinfo {pages} {e2221736120} (\bibinfo {year}
  {2023})}\BibitemShut {NoStop}%
\bibitem [{\citenamefont {Chou}\ \emph {et~al.}(2023)\citenamefont {Chou},
  \citenamefont {Shemma}, \citenamefont {McCarrick}, \citenamefont {Chien},
  \citenamefont {Teoh}, \citenamefont {Winkel}, \citenamefont {Anderson},
  \citenamefont {Chen}, \citenamefont {Curtis}, \citenamefont {de~Graaf} \emph
  {et~al.}}]{Chou2023Erasure}%
  \BibitemOpen
  \bibfield  {author} {\bibinfo {author} {\bibfnamefont {K.~S.}\ \bibnamefont
  {Chou}}, \bibinfo {author} {\bibfnamefont {T.}~\bibnamefont {Shemma}},
  \bibinfo {author} {\bibfnamefont {H.}~\bibnamefont {McCarrick}}, \bibinfo
  {author} {\bibfnamefont {T.-C.}\ \bibnamefont {Chien}}, \bibinfo {author}
  {\bibfnamefont {J.~D.}\ \bibnamefont {Teoh}}, \bibinfo {author}
  {\bibfnamefont {P.}~\bibnamefont {Winkel}}, \bibinfo {author} {\bibfnamefont
  {A.}~\bibnamefont {Anderson}}, \bibinfo {author} {\bibfnamefont
  {J.}~\bibnamefont {Chen}}, \bibinfo {author} {\bibfnamefont {J.}~\bibnamefont
  {Curtis}}, \bibinfo {author} {\bibfnamefont {S.~J.}\ \bibnamefont
  {de~Graaf}},  \emph {et~al.},\ }\href@noop {} {\bibfield  {journal} {\bibinfo
   {journal} {arXiv:2307.03169}\ } (\bibinfo {year} {2023})}\BibitemShut
  {NoStop}%
\bibitem [{\citenamefont {Levine}\ \emph {et~al.}(2024)\citenamefont {Levine},
  \citenamefont {Haim}, \citenamefont {Hung}, \citenamefont {Alidoust},
  \citenamefont {Kalaee}, \citenamefont {DeLorenzo}, \citenamefont {Wollack},
  \citenamefont {Arrangoiz-Arriola}, \citenamefont {Khalajhedayati},
  \citenamefont {Sanil}, \citenamefont {Moradinejad}, \citenamefont {Vaknin},
  \citenamefont {Kubica}, \citenamefont {Hover}, \citenamefont {Aghaeimeibodi},
  \citenamefont {Alcid}, \citenamefont {Baek}, \citenamefont {Barnett},
  \citenamefont {Bawdekar}, \citenamefont {Bienias}, \citenamefont {Carson},
  \citenamefont {Chen}, \citenamefont {Chen}, \citenamefont {Chinkezian},
  \citenamefont {Chisholm}, \citenamefont {Clifford}, \citenamefont {Cosmic},
  \citenamefont {Crisosto}, \citenamefont {Dalzell}, \citenamefont {Davis},
  \citenamefont {D'Ewart}, \citenamefont {Diez}, \citenamefont {D'Souza},
  \citenamefont {Dumitrescu}, \citenamefont {Elkhouly}, \citenamefont {Fang},
  \citenamefont {Fang}, \citenamefont {Flammia}, \citenamefont {Fling},
  \citenamefont {Garcia}, \citenamefont {Gharzai}, \citenamefont {Gorshkov},
  \citenamefont {Gray}, \citenamefont {Grimberg}, \citenamefont {Grimsmo},
  \citenamefont {Hann}, \citenamefont {He}, \citenamefont {Heidel},
  \citenamefont {Howell}, \citenamefont {Hunt}, \citenamefont {Iverson},
  \citenamefont {Jarrige}, \citenamefont {Jiang}, \citenamefont {Jones},
  \citenamefont {Karabalin}, \citenamefont {Karalekas}, \citenamefont {Keller},
  \citenamefont {Lasi}, \citenamefont {Lee}, \citenamefont {Ly}, \citenamefont
  {MacCabe}, \citenamefont {Mahuli}, \citenamefont {Marcaud}, \citenamefont
  {Matheny}, \citenamefont {McArdle}, \citenamefont {McCabe}, \citenamefont
  {Merton}, \citenamefont {Miles}, \citenamefont {Milsted}, \citenamefont
  {Mishra}, \citenamefont {Moncelsi}, \citenamefont {Naghiloo}, \citenamefont
  {Noh}, \citenamefont {Oblepias}, \citenamefont {Ortuno}, \citenamefont
  {Owens}, \citenamefont {Pagdilao}, \citenamefont {Panduro}, \citenamefont
  {Paquette}, \citenamefont {Patel}, \citenamefont {Peairs}, \citenamefont
  {Perello}, \citenamefont {Peterson}, \citenamefont {Ponte}, \citenamefont
  {Putterman}, \citenamefont {Refael}, \citenamefont {Reinhold}, \citenamefont
  {Resnick}, \citenamefont {Reyna}, \citenamefont {Rodriguez}, \citenamefont
  {Rose}, \citenamefont {Rubin}, \citenamefont {Runyan}, \citenamefont {Ryan},
  \citenamefont {Sahmoud}, \citenamefont {Scaffidi}, \citenamefont {Shah},
  \citenamefont {Siavoshi}, \citenamefont {Sivarajah}, \citenamefont
  {Skogland}, \citenamefont {Su}, \citenamefont {Swenson}, \citenamefont
  {Sylvia}, \citenamefont {Teo}, \citenamefont {Tomada}, \citenamefont
  {Torlai}, \citenamefont {Wistrom}, \citenamefont {Zhang}, \citenamefont
  {Zuk}, \citenamefont {Clerk}, \citenamefont {Brand\~ao}, \citenamefont
  {Retzker},\ and\ \citenamefont {Painter}}]{Levine2024Erasure}%
  \BibitemOpen
  \bibfield  {author} {\bibinfo {author} {\bibfnamefont {H.}~\bibnamefont
  {Levine}}, \bibinfo {author} {\bibfnamefont {A.}~\bibnamefont {Haim}},
  \bibinfo {author} {\bibfnamefont {J.~S.~C.}\ \bibnamefont {Hung}}, \bibinfo
  {author} {\bibfnamefont {N.}~\bibnamefont {Alidoust}}, \bibinfo {author}
  {\bibfnamefont {M.}~\bibnamefont {Kalaee}}, \bibinfo {author} {\bibfnamefont
  {L.}~\bibnamefont {DeLorenzo}}, \bibinfo {author} {\bibfnamefont {E.~A.}\
  \bibnamefont {Wollack}}, \bibinfo {author} {\bibfnamefont {P.}~\bibnamefont
  {Arrangoiz-Arriola}}, \bibinfo {author} {\bibfnamefont {A.}~\bibnamefont
  {Khalajhedayati}}, \bibinfo {author} {\bibfnamefont {R.}~\bibnamefont
  {Sanil}}, \bibinfo {author} {\bibfnamefont {H.}~\bibnamefont {Moradinejad}},
  \bibinfo {author} {\bibfnamefont {Y.}~\bibnamefont {Vaknin}}, \bibinfo
  {author} {\bibfnamefont {A.}~\bibnamefont {Kubica}}, \bibinfo {author}
  {\bibfnamefont {D.}~\bibnamefont {Hover}}, \bibinfo {author} {\bibfnamefont
  {S.}~\bibnamefont {Aghaeimeibodi}}, \bibinfo {author} {\bibfnamefont {J.~A.}\
  \bibnamefont {Alcid}}, \bibinfo {author} {\bibfnamefont {C.}~\bibnamefont
  {Baek}}, \bibinfo {author} {\bibfnamefont {J.}~\bibnamefont {Barnett}},
  \bibinfo {author} {\bibfnamefont {K.}~\bibnamefont {Bawdekar}}, \bibinfo
  {author} {\bibfnamefont {P.}~\bibnamefont {Bienias}}, \bibinfo {author}
  {\bibfnamefont {H.~A.}\ \bibnamefont {Carson}}, \bibinfo {author}
  {\bibfnamefont {C.}~\bibnamefont {Chen}}, \bibinfo {author} {\bibfnamefont
  {L.}~\bibnamefont {Chen}}, \bibinfo {author} {\bibfnamefont {H.}~\bibnamefont
  {Chinkezian}}, \bibinfo {author} {\bibfnamefont {E.~M.}\ \bibnamefont
  {Chisholm}}, \bibinfo {author} {\bibfnamefont {A.}~\bibnamefont {Clifford}},
  \bibinfo {author} {\bibfnamefont {R.}~\bibnamefont {Cosmic}}, \bibinfo
  {author} {\bibfnamefont {N.}~\bibnamefont {Crisosto}}, \bibinfo {author}
  {\bibfnamefont {A.~M.}\ \bibnamefont {Dalzell}}, \bibinfo {author}
  {\bibfnamefont {E.}~\bibnamefont {Davis}}, \bibinfo {author} {\bibfnamefont
  {J.~M.}\ \bibnamefont {D'Ewart}}, \bibinfo {author} {\bibfnamefont
  {S.}~\bibnamefont {Diez}}, \bibinfo {author} {\bibfnamefont {N.}~\bibnamefont
  {D'Souza}}, \bibinfo {author} {\bibfnamefont {P.~T.}\ \bibnamefont
  {Dumitrescu}}, \bibinfo {author} {\bibfnamefont {E.}~\bibnamefont
  {Elkhouly}}, \bibinfo {author} {\bibfnamefont {M.~T.}\ \bibnamefont {Fang}},
  \bibinfo {author} {\bibfnamefont {Y.}~\bibnamefont {Fang}}, \bibinfo {author}
  {\bibfnamefont {S.}~\bibnamefont {Flammia}}, \bibinfo {author} {\bibfnamefont
  {M.~J.}\ \bibnamefont {Fling}}, \bibinfo {author} {\bibfnamefont
  {G.}~\bibnamefont {Garcia}}, \bibinfo {author} {\bibfnamefont {M.~K.}\
  \bibnamefont {Gharzai}}, \bibinfo {author} {\bibfnamefont {A.~V.}\
  \bibnamefont {Gorshkov}}, \bibinfo {author} {\bibfnamefont {M.~J.}\
  \bibnamefont {Gray}}, \bibinfo {author} {\bibfnamefont {S.}~\bibnamefont
  {Grimberg}}, \bibinfo {author} {\bibfnamefont {A.~L.}\ \bibnamefont
  {Grimsmo}}, \bibinfo {author} {\bibfnamefont {C.~T.}\ \bibnamefont {Hann}},
  \bibinfo {author} {\bibfnamefont {Y.}~\bibnamefont {He}}, \bibinfo {author}
  {\bibfnamefont {S.}~\bibnamefont {Heidel}}, \bibinfo {author} {\bibfnamefont
  {S.}~\bibnamefont {Howell}}, \bibinfo {author} {\bibfnamefont
  {M.}~\bibnamefont {Hunt}}, \bibinfo {author} {\bibfnamefont {J.}~\bibnamefont
  {Iverson}}, \bibinfo {author} {\bibfnamefont {I.}~\bibnamefont {Jarrige}},
  \bibinfo {author} {\bibfnamefont {L.}~\bibnamefont {Jiang}}, \bibinfo
  {author} {\bibfnamefont {W.~M.}\ \bibnamefont {Jones}}, \bibinfo {author}
  {\bibfnamefont {R.}~\bibnamefont {Karabalin}}, \bibinfo {author}
  {\bibfnamefont {P.~J.}\ \bibnamefont {Karalekas}}, \bibinfo {author}
  {\bibfnamefont {A.~J.}\ \bibnamefont {Keller}}, \bibinfo {author}
  {\bibfnamefont {D.}~\bibnamefont {Lasi}}, \bibinfo {author} {\bibfnamefont
  {M.}~\bibnamefont {Lee}}, \bibinfo {author} {\bibfnamefont {V.}~\bibnamefont
  {Ly}}, \bibinfo {author} {\bibfnamefont {G.}~\bibnamefont {MacCabe}},
  \bibinfo {author} {\bibfnamefont {N.}~\bibnamefont {Mahuli}}, \bibinfo
  {author} {\bibfnamefont {G.}~\bibnamefont {Marcaud}}, \bibinfo {author}
  {\bibfnamefont {M.~H.}\ \bibnamefont {Matheny}}, \bibinfo {author}
  {\bibfnamefont {S.}~\bibnamefont {McArdle}}, \bibinfo {author} {\bibfnamefont
  {G.}~\bibnamefont {McCabe}}, \bibinfo {author} {\bibfnamefont
  {G.}~\bibnamefont {Merton}}, \bibinfo {author} {\bibfnamefont
  {C.}~\bibnamefont {Miles}}, \bibinfo {author} {\bibfnamefont
  {A.}~\bibnamefont {Milsted}}, \bibinfo {author} {\bibfnamefont
  {A.}~\bibnamefont {Mishra}}, \bibinfo {author} {\bibfnamefont
  {L.}~\bibnamefont {Moncelsi}}, \bibinfo {author} {\bibfnamefont
  {M.}~\bibnamefont {Naghiloo}}, \bibinfo {author} {\bibfnamefont
  {K.}~\bibnamefont {Noh}}, \bibinfo {author} {\bibfnamefont {E.}~\bibnamefont
  {Oblepias}}, \bibinfo {author} {\bibfnamefont {G.}~\bibnamefont {Ortuno}},
  \bibinfo {author} {\bibfnamefont {J.~C.}\ \bibnamefont {Owens}}, \bibinfo
  {author} {\bibfnamefont {J.}~\bibnamefont {Pagdilao}}, \bibinfo {author}
  {\bibfnamefont {A.}~\bibnamefont {Panduro}}, \bibinfo {author} {\bibfnamefont
  {J.-P.}\ \bibnamefont {Paquette}}, \bibinfo {author} {\bibfnamefont {R.~N.}\
  \bibnamefont {Patel}}, \bibinfo {author} {\bibfnamefont {G.}~\bibnamefont
  {Peairs}}, \bibinfo {author} {\bibfnamefont {D.~J.}\ \bibnamefont {Perello}},
  \bibinfo {author} {\bibfnamefont {E.~C.}\ \bibnamefont {Peterson}}, \bibinfo
  {author} {\bibfnamefont {S.}~\bibnamefont {Ponte}}, \bibinfo {author}
  {\bibfnamefont {H.}~\bibnamefont {Putterman}}, \bibinfo {author}
  {\bibfnamefont {G.}~\bibnamefont {Refael}}, \bibinfo {author} {\bibfnamefont
  {P.}~\bibnamefont {Reinhold}}, \bibinfo {author} {\bibfnamefont
  {R.}~\bibnamefont {Resnick}}, \bibinfo {author} {\bibfnamefont {O.~A.}\
  \bibnamefont {Reyna}}, \bibinfo {author} {\bibfnamefont {R.}~\bibnamefont
  {Rodriguez}}, \bibinfo {author} {\bibfnamefont {J.}~\bibnamefont {Rose}},
  \bibinfo {author} {\bibfnamefont {A.~H.}\ \bibnamefont {Rubin}}, \bibinfo
  {author} {\bibfnamefont {M.}~\bibnamefont {Runyan}}, \bibinfo {author}
  {\bibfnamefont {C.~A.}\ \bibnamefont {Ryan}}, \bibinfo {author}
  {\bibfnamefont {A.}~\bibnamefont {Sahmoud}}, \bibinfo {author} {\bibfnamefont
  {T.}~\bibnamefont {Scaffidi}}, \bibinfo {author} {\bibfnamefont
  {B.}~\bibnamefont {Shah}}, \bibinfo {author} {\bibfnamefont {S.}~\bibnamefont
  {Siavoshi}}, \bibinfo {author} {\bibfnamefont {P.}~\bibnamefont {Sivarajah}},
  \bibinfo {author} {\bibfnamefont {T.}~\bibnamefont {Skogland}}, \bibinfo
  {author} {\bibfnamefont {C.-J.}\ \bibnamefont {Su}}, \bibinfo {author}
  {\bibfnamefont {L.~J.}\ \bibnamefont {Swenson}}, \bibinfo {author}
  {\bibfnamefont {J.}~\bibnamefont {Sylvia}}, \bibinfo {author} {\bibfnamefont
  {S.~M.}\ \bibnamefont {Teo}}, \bibinfo {author} {\bibfnamefont
  {A.}~\bibnamefont {Tomada}}, \bibinfo {author} {\bibfnamefont
  {G.}~\bibnamefont {Torlai}}, \bibinfo {author} {\bibfnamefont
  {M.}~\bibnamefont {Wistrom}}, \bibinfo {author} {\bibfnamefont
  {K.}~\bibnamefont {Zhang}}, \bibinfo {author} {\bibfnamefont
  {I.}~\bibnamefont {Zuk}}, \bibinfo {author} {\bibfnamefont {A.~A.}\
  \bibnamefont {Clerk}}, \bibinfo {author} {\bibfnamefont {F.~G. S.~L.}\
  \bibnamefont {Brand\~ao}}, \bibinfo {author} {\bibfnamefont {A.}~\bibnamefont
  {Retzker}}, \ and\ \bibinfo {author} {\bibfnamefont {O.}~\bibnamefont
  {Painter}},\ }\href {\doibase 10.1103/PhysRevX.14.011051} {\bibfield
  {journal} {\bibinfo  {journal} {Phys. Rev. X}\ }\textbf {\bibinfo {volume}
  {14}},\ \bibinfo {pages} {011051} (\bibinfo {year} {2024})}\BibitemShut
  {NoStop}%
\bibitem [{\citenamefont {Hutzler}\ \emph {et~al.}(2012)\citenamefont
  {Hutzler}, \citenamefont {Lu},\ and\ \citenamefont
  {Doyle}}]{Hutzler2012CBGB}%
  \BibitemOpen
  \bibfield  {author} {\bibinfo {author} {\bibfnamefont {N.~R.}\ \bibnamefont
  {Hutzler}}, \bibinfo {author} {\bibfnamefont {H.-I.}\ \bibnamefont {Lu}}, \
  and\ \bibinfo {author} {\bibfnamefont {J.~M.}\ \bibnamefont {Doyle}},\ }\href
  {https://doi.org/10.1021/cr200362u} {\bibfield  {journal} {\bibinfo
  {journal} {Chem. Rev.}\ }\textbf {\bibinfo {volume} {112}},\ \bibinfo {pages}
  {4803} (\bibinfo {year} {2012})}\BibitemShut {NoStop}%
\bibitem [{\citenamefont {Truppe}\ \emph
  {et~al.}(2017{\natexlab{a}})\citenamefont {Truppe}, \citenamefont {Williams},
  \citenamefont {Fitch}, \citenamefont {Hambach}, \citenamefont {Wall},
  \citenamefont {Hinds}, \citenamefont {Sauer},\ and\ \citenamefont
  {Tarbutt}}]{Truppe2017chirp}%
  \BibitemOpen
  \bibfield  {author} {\bibinfo {author} {\bibfnamefont {S.}~\bibnamefont
  {Truppe}}, \bibinfo {author} {\bibfnamefont {H.~J.}\ \bibnamefont
  {Williams}}, \bibinfo {author} {\bibfnamefont {N.~J.}\ \bibnamefont {Fitch}},
  \bibinfo {author} {\bibfnamefont {M.}~\bibnamefont {Hambach}}, \bibinfo
  {author} {\bibfnamefont {T.~E.}\ \bibnamefont {Wall}}, \bibinfo {author}
  {\bibfnamefont {E.~A.}\ \bibnamefont {Hinds}}, \bibinfo {author}
  {\bibfnamefont {B.~E.}\ \bibnamefont {Sauer}}, \ and\ \bibinfo {author}
  {\bibfnamefont {M.~R.}\ \bibnamefont {Tarbutt}},\ }\href@noop {} {\bibfield
  {journal} {\bibinfo  {journal} {New Journal of Physics}\ }\textbf {\bibinfo
  {volume} {19}},\ \bibinfo {pages} {022001} (\bibinfo {year}
  {2017}{\natexlab{a}})}\BibitemShut {NoStop}%
\bibitem [{\citenamefont {Anderegg}\ \emph {et~al.}(2017)\citenamefont
  {Anderegg}, \citenamefont {Augenbraun}, \citenamefont {Chae}, \citenamefont
  {Hemmerling}, \citenamefont {Hutzler}, \citenamefont {Ravi}, \citenamefont
  {Collopy}, \citenamefont {Ye}, \citenamefont {Ketterle},\ and\ \citenamefont
  {Doyle}}]{Anderegg2017MOT}%
  \BibitemOpen
  \bibfield  {author} {\bibinfo {author} {\bibfnamefont {L.}~\bibnamefont
  {Anderegg}}, \bibinfo {author} {\bibfnamefont {B.~L.}\ \bibnamefont
  {Augenbraun}}, \bibinfo {author} {\bibfnamefont {E.}~\bibnamefont {Chae}},
  \bibinfo {author} {\bibfnamefont {B.}~\bibnamefont {Hemmerling}}, \bibinfo
  {author} {\bibfnamefont {N.~R.}\ \bibnamefont {Hutzler}}, \bibinfo {author}
  {\bibfnamefont {A.}~\bibnamefont {Ravi}}, \bibinfo {author} {\bibfnamefont
  {A.}~\bibnamefont {Collopy}}, \bibinfo {author} {\bibfnamefont
  {J.}~\bibnamefont {Ye}}, \bibinfo {author} {\bibfnamefont {W.}~\bibnamefont
  {Ketterle}}, \ and\ \bibinfo {author} {\bibfnamefont {J.~M.}\ \bibnamefont
  {Doyle}},\ }\href {\doibase 10.1103/PhysRevLett.119.103201} {\bibfield
  {journal} {\bibinfo  {journal} {Phys. Rev. Lett.}\ }\textbf {\bibinfo
  {volume} {119}},\ \bibinfo {pages} {103201} (\bibinfo {year}
  {2017})}\BibitemShut {NoStop}%
\bibitem [{\citenamefont {Truppe}\ \emph
  {et~al.}(2017{\natexlab{b}})\citenamefont {Truppe}, \citenamefont {Williams},
  \citenamefont {Hambach}, \citenamefont {Caldwell}, \citenamefont {Fitch},
  \citenamefont {Hinds}, \citenamefont {Sauer},\ and\ \citenamefont
  {Tarbutt}}]{Truppe2017Mot}%
  \BibitemOpen
  \bibfield  {author} {\bibinfo {author} {\bibfnamefont {S.}~\bibnamefont
  {Truppe}}, \bibinfo {author} {\bibfnamefont {H.~J.}\ \bibnamefont
  {Williams}}, \bibinfo {author} {\bibfnamefont {M.}~\bibnamefont {Hambach}},
  \bibinfo {author} {\bibfnamefont {L.}~\bibnamefont {Caldwell}}, \bibinfo
  {author} {\bibfnamefont {N.~J.}\ \bibnamefont {Fitch}}, \bibinfo {author}
  {\bibfnamefont {E.~A.}\ \bibnamefont {Hinds}}, \bibinfo {author}
  {\bibfnamefont {B.~E.}\ \bibnamefont {Sauer}}, \ and\ \bibinfo {author}
  {\bibfnamefont {M.~R.}\ \bibnamefont {Tarbutt}},\ }\href {\doibase
  10.1038/nphys4241} {\bibfield  {journal} {\bibinfo  {journal} {Nature
  Physics}\ }\textbf {\bibinfo {volume} {13}},\ \bibinfo {pages} {1173}
  (\bibinfo {year} {2017}{\natexlab{b}})}\BibitemShut {NoStop}%
\bibitem [{\citenamefont {Cheuk}\ \emph {et~al.}(2018)\citenamefont {Cheuk},
  \citenamefont {Anderegg}, \citenamefont {Augenbraun}, \citenamefont {Bao},
  \citenamefont {Burchesky}, \citenamefont {Ketterle},\ and\ \citenamefont
  {Doyle}}]{Cheuk2018Lambda}%
  \BibitemOpen
  \bibfield  {author} {\bibinfo {author} {\bibfnamefont {L.~W.}\ \bibnamefont
  {Cheuk}}, \bibinfo {author} {\bibfnamefont {L.}~\bibnamefont {Anderegg}},
  \bibinfo {author} {\bibfnamefont {B.~L.}\ \bibnamefont {Augenbraun}},
  \bibinfo {author} {\bibfnamefont {Y.}~\bibnamefont {Bao}}, \bibinfo {author}
  {\bibfnamefont {S.}~\bibnamefont {Burchesky}}, \bibinfo {author}
  {\bibfnamefont {W.}~\bibnamefont {Ketterle}}, \ and\ \bibinfo {author}
  {\bibfnamefont {J.~M.}\ \bibnamefont {Doyle}},\ }\href {\doibase
  10.1103/PhysRevLett.121.083201} {\bibfield  {journal} {\bibinfo  {journal}
  {Phys. Rev. Lett.}\ }\textbf {\bibinfo {volume} {121}},\ \bibinfo {pages}
  {083201} (\bibinfo {year} {2018})}\BibitemShut {NoStop}%
\bibitem [{\citenamefont {Li}\ \emph {et~al.}(2024)\citenamefont {Li},
  \citenamefont {Holland}, \citenamefont {Lu},\ and\ \citenamefont
  {Cheuk}}]{Li2023blueMOTCaF}%
  \BibitemOpen
  \bibfield  {author} {\bibinfo {author} {\bibfnamefont {S.~J.}\ \bibnamefont
  {Li}}, \bibinfo {author} {\bibfnamefont {C.~M.}\ \bibnamefont {Holland}},
  \bibinfo {author} {\bibfnamefont {Y.}~\bibnamefont {Lu}}, \ and\ \bibinfo
  {author} {\bibfnamefont {L.~W.}\ \bibnamefont {Cheuk}},\ }\href@noop {}
  {\bibfield  {journal} {\bibinfo  {journal} {Physical Review Letters}\ }
  (\bibinfo {year} {2024})},\ \Eprint {http://arxiv.org/abs/2311.05447v1}
  {2311.05447v1} \BibitemShut {NoStop}%
\bibitem [{\citenamefont {Anderegg}\ \emph {et~al.}(2018)\citenamefont
  {Anderegg}, \citenamefont {Augenbraun}, \citenamefont {Bao}, \citenamefont
  {Burchesky}, \citenamefont {Cheuk}, \citenamefont {Ketterle},\ and\
  \citenamefont {Doyle}}]{Anderegg2018ODT}%
  \BibitemOpen
  \bibfield  {author} {\bibinfo {author} {\bibfnamefont {L.}~\bibnamefont
  {Anderegg}}, \bibinfo {author} {\bibfnamefont {B.~L.}\ \bibnamefont
  {Augenbraun}}, \bibinfo {author} {\bibfnamefont {Y.}~\bibnamefont {Bao}},
  \bibinfo {author} {\bibfnamefont {S.}~\bibnamefont {Burchesky}}, \bibinfo
  {author} {\bibfnamefont {L.~W.}\ \bibnamefont {Cheuk}}, \bibinfo {author}
  {\bibfnamefont {W.}~\bibnamefont {Ketterle}}, \ and\ \bibinfo {author}
  {\bibfnamefont {J.~M.}\ \bibnamefont {Doyle}},\ }\href {\doibase
  10.1038/s41567-018-0191-z} {\bibfield  {journal} {\bibinfo  {journal} {Nature
  Physics}\ }\textbf {\bibinfo {volume} {14}},\ \bibinfo {pages} {890}
  (\bibinfo {year} {2018})}\BibitemShut {NoStop}%
\bibitem [{Sup()}]{Supplement}%
  \BibitemOpen
  \href@noop {} {}\bibinfo {note} {See Supplemental Material.}\BibitemShut
  {Stop}%
\bibitem [{\citenamefont {Hou}\ and\ \citenamefont
  {Bernath}(2018)}]{hou2018VibDip}%
  \BibitemOpen
  \bibfield  {author} {\bibinfo {author} {\bibfnamefont {S.}~\bibnamefont
  {Hou}}\ and\ \bibinfo {author} {\bibfnamefont {P.~F.}\ \bibnamefont
  {Bernath}},\ }\href@noop {} {\bibfield  {journal} {\bibinfo  {journal}
  {Journal of Quantitative Spectroscopy and Radiative Transfer}\ }\textbf
  {\bibinfo {volume} {210}},\ \bibinfo {pages} {44} (\bibinfo {year}
  {2018})}\BibitemShut {NoStop}%
\bibitem [{\citenamefont {Singh}\ \emph {et~al.}(2022)\citenamefont {Singh},
  \citenamefont {Anand}, \citenamefont {Pocklington}, \citenamefont {Kemp},\
  and\ \citenamefont {Bernien}}]{Singh2022RbCsAtomArray}%
  \BibitemOpen
  \bibfield  {author} {\bibinfo {author} {\bibfnamefont {K.}~\bibnamefont
  {Singh}}, \bibinfo {author} {\bibfnamefont {S.}~\bibnamefont {Anand}},
  \bibinfo {author} {\bibfnamefont {A.}~\bibnamefont {Pocklington}}, \bibinfo
  {author} {\bibfnamefont {J.~T.}\ \bibnamefont {Kemp}}, \ and\ \bibinfo
  {author} {\bibfnamefont {H.}~\bibnamefont {Bernien}},\ }\href
  {https://journals.aps.org/prx/abstract/10.1103/PhysRevX.12.011040} {\bibfield
   {journal} {\bibinfo  {journal} {Physical Review X}\ }\textbf {\bibinfo
  {volume} {12}},\ \bibinfo {pages} {011040} (\bibinfo {year}
  {2022})}\BibitemShut {NoStop}%
\bibitem [{\citenamefont {Shaw}\ \emph {et~al.}(2023)\citenamefont {Shaw},
  \citenamefont {Scholl}, \citenamefont {Finklestein}, \citenamefont
  {Madjarov}, \citenamefont {Grinkemeyer},\ and\ \citenamefont
  {Endres}}]{Shaw2023DarkStateLoading}%
  \BibitemOpen
  \bibfield  {author} {\bibinfo {author} {\bibfnamefont {A.~L.}\ \bibnamefont
  {Shaw}}, \bibinfo {author} {\bibfnamefont {P.}~\bibnamefont {Scholl}},
  \bibinfo {author} {\bibfnamefont {R.}~\bibnamefont {Finklestein}}, \bibinfo
  {author} {\bibfnamefont {I.~S.}\ \bibnamefont {Madjarov}}, \bibinfo {author}
  {\bibfnamefont {B.}~\bibnamefont {Grinkemeyer}}, \ and\ \bibinfo {author}
  {\bibfnamefont {M.}~\bibnamefont {Endres}},\ }\href
  {https://journals.aps.org/prl/abstract/10.1103/PhysRevLett.130.193402}
  {\bibfield  {journal} {\bibinfo  {journal} {Physical Review Letters}\
  }\textbf {\bibinfo {volume} {130}},\ \bibinfo {pages} {193402} (\bibinfo
  {year} {2023})}\BibitemShut {NoStop}%
\bibitem [{\citenamefont {Homeier}\ \emph {et~al.}(2023)\citenamefont
  {Homeier}, \citenamefont {Harris}, \citenamefont {Blatz}, \citenamefont
  {Schollw{\"o}ck}, \citenamefont {Grusdt},\ and\ \citenamefont
  {Bohrdt}}]{homeier2023AFMtJ}%
  \BibitemOpen
  \bibfield  {author} {\bibinfo {author} {\bibfnamefont {L.}~\bibnamefont
  {Homeier}}, \bibinfo {author} {\bibfnamefont {T.~J.}\ \bibnamefont {Harris}},
  \bibinfo {author} {\bibfnamefont {T.}~\bibnamefont {Blatz}}, \bibinfo
  {author} {\bibfnamefont {U.}~\bibnamefont {Schollw{\"o}ck}}, \bibinfo
  {author} {\bibfnamefont {F.}~\bibnamefont {Grusdt}}, \ and\ \bibinfo {author}
  {\bibfnamefont {A.}~\bibnamefont {Bohrdt}},\ }\href@noop {} {\bibfield
  {journal} {\bibinfo  {journal} {arXiv preprint arXiv:2305.02322}\ } (\bibinfo
  {year} {2023})}\BibitemShut {NoStop}%
\bibitem [{\citenamefont {Halimeh}\ \emph {et~al.}(2023)\citenamefont
  {Halimeh}, \citenamefont {Homeier}, \citenamefont {Bohrdt},\ and\
  \citenamefont {Grusdt}}]{Halimeh2023Rishon}%
  \BibitemOpen
  \bibfield  {author} {\bibinfo {author} {\bibfnamefont {J.~C.}\ \bibnamefont
  {Halimeh}}, \bibinfo {author} {\bibfnamefont {L.}~\bibnamefont {Homeier}},
  \bibinfo {author} {\bibfnamefont {A.}~\bibnamefont {Bohrdt}}, \ and\ \bibinfo
  {author} {\bibfnamefont {F.}~\bibnamefont {Grusdt}},\ }\href@noop {}
  {\bibfield  {journal} {\bibinfo  {journal} {arXiv preprint arXiv:2305.06373}\
  } (\bibinfo {year} {2023})}\BibitemShut {NoStop}%
\bibitem [{\citenamefont {Norrgard}\ \emph {et~al.}(2021)\citenamefont
  {Norrgard}, \citenamefont {Eckel}, \citenamefont {Holloway},\ and\
  \citenamefont {Shirley}}]{Norrgard2021QBBTherm}%
  \BibitemOpen
  \bibfield  {author} {\bibinfo {author} {\bibfnamefont {E.~B.}\ \bibnamefont
  {Norrgard}}, \bibinfo {author} {\bibfnamefont {S.~P.}\ \bibnamefont {Eckel}},
  \bibinfo {author} {\bibfnamefont {C.~L.}\ \bibnamefont {Holloway}}, \ and\
  \bibinfo {author} {\bibfnamefont {E.~L.}\ \bibnamefont {Shirley}},\ }\href
  {https://iopscience.iop.org/article/10.1088/1367-2630/abe8f5/meta} {\bibfield
   {journal} {\bibinfo  {journal} {New Journal of Physics}\ }\textbf {\bibinfo
  {volume} {23}},\ \bibinfo {pages} {033037} (\bibinfo {year}
  {2021})}\BibitemShut {NoStop}%
\end{thebibliography}

\end{document}